\documentclass[prb,twocolumn,showpacs,amsmath,amssymb,superscriptaddress]{revtex4}

\usepackage{graphicx}
\usepackage{color}
\usepackage{bbm}

\date{\today}                  

\def\ea{\emph{et al.}}

\begin{document}

\title{Dissipative quantum mechanics beyond Bloch-Redfield: A consistent
  weak-coupling expansion of the ohmic spin boson model at arbitrary bias}

\author{Carsten J. Lindner}
\affiliation{Institut f\"ur Theorie der Statistischen Physik, RWTH Aachen, 
52056 Aachen, Germany and JARA - Fundamentals of Future Information Technology}
\author{Herbert Schoeller}
\email[Email: ]{schoeller@physik.rwth-aachen.de}
\affiliation{Institut f\"ur Theorie der Statistischen Physik, RWTH Aachen, 
52056 Aachen, Germany and JARA - Fundamentals of Future Information Technology}

\begin{abstract}
The ohmic spin boson model at finite bias $\epsilon$ and tunneling $\Delta$ is an 
important model to study relaxation and decoherence properties of qubits coupled weakly 
to a dissipative bosonic environment. 
Fault-tolerant quantum computation requires the investigation of small errors beyond 
Bloch-Redfield theory. Using perturbation theory and the real-time renormalization group (RG) method 
we present a consistent zero-temperature weak-coupling expansion for the
time evolution of the reduced density matrix $\rho(t)$ out of an uncorrelated but otherwise arbitrary initial 
state. We show that standard Born approximation schemes calulating the effective Liouvillian 
of the kinetic equation up to first order in the dimensionless coupling constant $\alpha$ 
are not sufficient to account for various important corrections one order beyond the Bloch-Redfield solution:
(1) The resummation of {\it all} secular terms $\sim(\Gamma t)^n$ is necessary to obtain the correct 
exponential decay of {\it all} terms of the time evolution with decay rate $\Gamma$ or $\Gamma/2$, 
together with the correct 
preexponential functions. We show that this is only possible by a correct analytic continuation 
of the Fourier transform $L(E)$ of the effective Liouvillian into the lower half of the
complex plane. (2) The resummation of all logarithmic terms at high and low energies leads to a 
renormalized tunneling $\tilde{\Delta}$ and to preexponential functions 
of logarithmic and power-law form. This is achieved by solving closed differential equations for the
derivatives of $L(E)$ set up by the real-time RG method. (3) The fact that two eigenvalues of $L(E)$ 
are close to each other by $O(\Gamma)$ requires degenerate perturbation theory for times $\Gamma t\sim O(1)$,
where certain terms of the Liouvillian in $O(\alpha^2)$ are needed to calculate the 
stationary state and the time evolution of the non-oscillating purely decaying terms up to $O(\alpha)$. 
Solving the real-time RG equations perturbatively we present a renormalized perturbation theory
with analytical results for arbitrary bias covering the whole crossover regime from small times 
$\Omega t\ll 1$ to large times $\Omega t\gg 1$, where $\Omega=\sqrt{\epsilon^2+\tilde{\Delta}^2}$ denotes 
the renormalized level splitting, and compare to other results published in the literature. We
find that branch cuts starting at pole positions lead to rather slowly varying logarithmic time
dependencies for the preexponential functions, whereas pure branch cuts give rise to a strong
time dependence of the preexponential functions in the crossover regime with a leading long time tail 
$\sim \alpha/(\Omega t)$ at finite bias, besides other terms $\sim \alpha/(\Omega t)^2$ well-known from 
the zero-bias case. For exponentially large times, we use the real-time RG method for a 
non-perturbative resummation of all logarithmic terms $\sim (\alpha\ln(\Omega t))^n$  and present
a numerical solution for arbitrary bias. 
We find an interesting power-law behaviour $({1\over \Omega t})^{2\alpha{\epsilon^2\over \Omega^2}}$ 
with a bias-dependent power-law exponent appearing in the preexponential functions of the oscillating terms.
This power-law has to be contrasted to the one at exponentially small times where the power-law exponent
crosses over to $2\alpha$. We discuss that the complexity to calculate one order beyond Bloch-Redfield 
is rather generic and applies also to other models of dissipative quantum mechanics. 

\end{abstract}


\maketitle

\section{Introduction}
\label{sec:introduction}

The study of the dynamics of two-state quantum systems coupled weakly to a dissipative bath is
a fundamental problem of nonequilibrium statistical mechanics that has become of increasing 
importance due to possible future technological applications in quantum information
processing. To realize scalable and fault-tolerant quantum computation very low error
thresholds are needed which requires an understanding beyond Markov approximation schemes and
lowest order perturbation theory in the coupling to the bath. As a generic model for a bosonic bath 
the spin-boson model has been proposed \cite{leggett_87} and its dynamical properties have been studied
with various methods \cite{weiss_12,grifoni_haenggi_98}. This model consists of two levels with level
spacing (bias) $\epsilon$, coupled by a direct tunneling term $\Delta$, and each level is linearly coupled  
to an ohmic bosonic bath. In the case of zero tunneling, the spin-boson model can be solved exactly 
and the stability of
surface-code error correction against realistic dissipation has been studied recently 
\cite{hutter_loss_13,novais_mucciolo_13}. However, for finite tunneling and the most important case 
of an ohmic coupling to the bath, we will show in this work that a consistent weak-coupling expansion 
beyond the Bloch-Redfield Markov approximation is still lacking at low temperatures. We will discuss 
various subtleties to obtain a consistent perturbative expansion of the time evolution in the
dimensionless coupling constant $\alpha$, requiring an essentially non-perturbative treatment in a 
certain sense, not yet accounted for completely in various previous publications on the ohmic spin boson
model. Most importantly, our analysis shows that the systematic calculation of errors to the 
Bloch-Redfield result is generically very complex for all models of dissipative quantum mechanics, 
involving many details of the underlying model, and is not specific to the ohmic spin boson model.

The concrete form of the time evolution depends crucially on the form of the density of states 
of the bath and the energy dependence of the coupling constants $g_q$ of the local system to the 
various bath modes $\omega_q$, conveniently taken together in the spectral density (sometimes also 
called hybridization function) $J(\omega)\sim\sum_q g_q^2\delta(\omega-\omega_q)$. However, for a 
flat spectral density (on the scale of the typical energy scales of the local system) or for special 
cases as the ohmic spin boson model, where $J(\omega)\sim\alpha\omega$, a rather generic discussion 
of the typical form of the time evolution is possible and has been provided in 
Ref.~\onlinecite{pletyukhov_schuricht_schoeller_10}, based on the real-time renormalization group method 
(RTRG), see Rfs.~\onlinecite{schoeller_09,schoeller_14} for reviews. Starting at time $t=0$ from an arbitrary 
initial state $\rho_0=\rho(t=0)$ of the local system without any initial system-bath correlations 
(i.e. the bath is assumed 
to be an infinitely large system in (grand)canonical equilibrium), the time evolution for the reduced density
matrix $\rho(t)$ of the local system consists of a sum of terms each of them being $\sim {\cal{F}}_i(t)e^{-iz_it}$,
with $z_i=\Omega_i-i\Gamma_i$, i.e. exponentially decaying with decay rate $\Gamma_i>0$, oscillating 
with frequency $\Omega_i$, together with a non-exponential preexponential function ${\cal{F}}_i(t)$, typically 
depending logarithmically or as some power-law on time. The case $\Gamma_i=0$ is exceptional and occurs only
for systems with quantum critical points, where the scaling behavior is not cut off by any decay rate. 
For the ohmic spin boson model there are three modes of a purely decaying mode $z_0=-i\Gamma$ and two
oscillating modes with $z_\pm=\pm\Omega-i\Gamma/2$. 
This form already suggests where the complexity of calculating the time evolution beyond the lowest
order Markovian Bloch-Redfield theory appears. Bloch-Redfield considers only the leading order term
where ${\cal{F}}_i$ is basically a constant of $O(1)$, independent of the coupling $\alpha$ to the bath.
However, there are additional terms to each matrix element of the $2\times 2$-matrix $\rho(t)$, where 
${\cal{F}}_i(t)\sim\alpha$, also containing an exponential function, usually different from the one of the
Bloch-Redfield term. Expanding this exponential in $\Gamma\sim\alpha$ leads to an ill-defined 
perturbation expansion, since terms $\sim\alpha(\Gamma t)^n$ appear, which all become of $O(\alpha)$ already on 
time scales of the inverse decay rate (and even diverge for time going to infinity). Therefore, for a 
consistent calculation of the $O(\alpha)$ correction to Bloch-Redfield on time scales where the exponential
damping is still moderate, it is necessary to resum these terms in all orders of perturbation theory to get
the correct exponential behavior. We note that these so-called secular terms (sometimes also called 
van Hove singularities \cite{van_hove_loss}) are usually only discussed when expanding   
the exponentials of the Bloch-Redfield terms in $\alpha$, but similarly also appear in higher
order terms, which are more subtle. Technically, they can all be incorporated by expressing the
perturbative expansion for the effective Liouvillian $L(E)$ in Fourier space not in terms of the
bare Liouvillian but in terms of the full Liouvillian again by taking all self-energy insertions into
account. Within the diagrammatic expansion developed in 
Rfs.~\onlinecite{schoeller_09,schoeller_14,pletyukhov_schoeller_12,kashuba_schoeller_13} it can
be seen that this is possible in all orders of perturbation theory. This allows for a convenient 
analytic continuation of $L(E)$ into the lower half of the complex plane, from which the position
of all non-zero singularities $z_i$ (poles and branching points) of the Fourier transform 
$\rho(E)=i/(E-L(E))\rho_0$ can be determined self-consistently, leading to the effect that all
$z_i$ acquire a finite imaginary part $-i\Gamma_i$.

In connection with the ohmic spin boson model at zero bias, the 
occurrence of exponentials in the $O(\alpha)$-correction to Bloch-Redfield has recently been noted and 
corrected in Rfs.~\onlinecite{slutskin_etal_11,kashuba_schoeller_13}. Similar considerations have been performed  
close to $\alpha\sim{1\over 2}$, see Rfs.~\onlinecite{egger_etal_97,kennes_etal_13}. For finite bias, 
a Born approximation has been used in Ref.~\onlinecite{divincenzo_loss_05} to calculate perturbatively 
one order beyond Bloch Redfield, missing the exponentials in those corrections. In this paper 
we will present a perturbative calculation at arbitrary bias including all exponentials and, moreover, 
show that the resummation of secular terms is also important to obtain the correct energy scales in
logarithmic terms of preexponential functions. Furthermore, we will calculate all terms of the
time evolution for an arbitrary initial state of the local system, whereas in 
Ref.~\onlinecite{divincenzo_loss_05} only the time evolution of the Pauli matrix in $z$-direction has
been calculated for an initial state without any spin in $x-$ and $y-$ direction.

Besides secular terms proportional to powers of time, there are further subtleties in the calculation
of the time evolution, even in the case where potential logarithmic terms can be treated perturbatively. 
A generic feature of the reduced density matrix $\rho(E)=i/(E-L(E))\rho_0$ in Fourier space is that there
occurs one singularity at $E=0$ (determining the stationary state from $L(0^+)\rho_{\text{st}}=0$) and
a pure decay pole at $E=z_0=-i\Gamma\sim O(\alpha)$. These two singularities are close to each other 
within the expansion parameter $\alpha$, and leads to the generic feature that two eigenvalues of 
$L(E)$ are close to each other by $O(\alpha)$. Therefore, degenerate perturbation theory is necessary
for the zero and purely decaying modes, and the calculation of the corresponding projectors on the 
eigenstates of $L(E)$ up to $O(\alpha)$ requires the knowledge of the Liouvillian at least up to 
$O(\alpha^2)$. This fact has already been mentioned at the end of Ref.~\onlinecite{divincenzo_loss_05}, where the 
stationary state was calculated up to $O(\alpha)$ and the influence on the time evolution for the
purely decaying mode was indicated. This again is a generic problem for all models of dissipative
quantum mechanics and shows that lowest order Born approximation is not sufficient to account for all
first-order corrections to Bloch-Redfield. 
In this paper we will show that the special algebra of the ohmic spin model allows for
a simplification of this problem such that the results of 
Ref.~\onlinecite{divincenzo_loss_05} for the stationary case up to $O(\alpha)$ can be used to calculate also
all terms in $O(\alpha)$ for the time evolution of the purely decaying mode. 

The ohmic spin boson model (and similar many other models with a rather structureless spectral density of
states) has further problems in perturbation theory arising from logarithmically divergent integrals at
high and low energies, which have to be treated by renormalization group. At high energies logarithmic 
divergencies $\sim\alpha\ln{D/\Lambda_c}$ occur, where $D$ denotes the finite band width and
$\Lambda_c\sim\text{max}\{1/t,\Omega\}$ is some high-energy cutoff determined by the largest energy scale 
of the system. For large $D$ a non-perturbative resummation
of all powers of such terms is required. In the short-time regime $1/t\gg\Omega$, this leads to 
well-known terms $\sim 1/(D t)^{2\alpha}$ which can also be obtained from
the noninteracting blip approximation (NIBA) \cite{leggett_87,weiss_12}. For the most important
regime of times which are not exponentially small or large, where $|\alpha\ln(\Omega t)|\ll 1$, 
we will show in this paper that the logarithmic terms at high energies can be 
incorporated into a renormalized tunneling $\tilde{\Delta}=\Delta(\Omega/D)^\alpha$, where 
$\Omega=\sqrt{\epsilon^2+\tilde{\Delta}^2}$ is the renormalized Rabi frequency of the local system,
leading also to a renormalized decay rate $\Gamma=\pi\alpha\tilde{\Delta}^2/\Omega$. We note that the
correct cutoff scale is $\Lambda_c=\Omega$ and {\it not} $\tilde{\Delta}$ as first pointed out 
in Ref.~\onlinecite{divincenzo_loss_05},
where the logarithmic correction was calculated perturbatively in $\alpha$. Furthermore, we
will show in this paper how the unrenormalized tunneling occurring in various terms of perturbation 
theory has to be replaced by the renormalized one. This is quite nontrivial since both $\Delta$
and $\tilde{\Delta}$ appear in the final solution. We will achieve this goal by solving the RTRG equations
perturbatively with the result of a renormalized propagator containing $Z$-factors with 
$Z=\tilde{\Delta}^2/\Delta^2$. Subsequently, we will apply renormalized perturbation theory to calculate 
the time evolution analytically in the whole crossover regime from small times $\Omega t~\ll 1$
to large times $\Omega t\gg 1$ with $|\alpha\ln(\Omega t)|\ll 1$ such that logarithmic terms in time can 
be treated perturbatively. We find that the leading order terms in the preexponential functions
stem from branch cuts starting at a pole position of $\rho(E)$ giving rise to constant terms together with terms 
$\sim\alpha\ln(\Omega t)$ showing a rather weak logarithmic time dependence. In contrast, branch cuts 
starting at branching points unequal to the poles of $\rho(E)$ lead to crossover functions with a
strong time dependence of the preexponential functions which all can be expressed by the exponential integral. 
Interestingly, for large times $\Omega t\gg 1$, we find that the leading order terms fall off 
$\sim \alpha/(\Omega t)$ for finite bias, in contrast to the unbiased case, where all terms fall off 
$\sim \alpha/(\Omega t)^2$.  

After having got rid of logarithmic terms at high
energies, one is still left with logarithmic terms at low energies $\sim\alpha\ln{\Omega t}$. If the latter 
can be treated perturbatively, the solution for the time evolution one order beyond Bloch-Redfield  
follows from the above mentioned renormalized perturbation theory with the proper replacement of 
$\Delta$ by $\tilde{\Delta}$. 
However, for intermediate couplings $\alpha\sim 0.1-0.2$ or for the case of high bias 
$\epsilon\gg\tilde{\Delta}$ (where the decay rate $\Gamma\ll\alpha\Omega$ is very small), it turns out 
that higher order terms $\sim(\alpha\ln{\Omega t})^n$ with $n>1$ become important already for times
scales $t\sim 1/\Gamma$. In these cases a non-perturbative resummation is also necessary for the logarithmic 
terms at low energies to determine the first correction to Bloch-Redfield consistently. The only 
available method up to date to achieve such a resummation is the RTRG method 
\cite{schoeller_09,schoeller_14,pletyukhov_schoeller_12,kashuba_schoeller_13},
which can account simultaneously for logarithmic terms at high and low energies in all orders 
to determine the time evolution of models of dissipative quantum mechanics in the weak coupling regime. 
The idea is not to consider the perturbative expansion of the effective Liouvillian $L(E)$ but of 
the second derivative ${\text{d}^2\over \text{d}E^2}L(E)$, 
together with a proper resummation of self-energy insertions
and vertex corrections. This leads to a set of closed differential equations for the effective 
Liouvillian and the effective vertices, which are well-defined in the limit $D\rightarrow\infty$ and contain 
no secular terms and logarithmic divergencies at low and high energies. 
Therefore, the r.h.s. of these differential equations are a well-defined series in $\alpha$ and can be 
truncated systematically. We will consider the RG equations in leading order and solve them 
numerically for the ohmic spin boson model at arbitrary bias. Most importantly, we find for the
leading order terms of the preexponential functions of the oscillating modes a power-law behaviour 
$\sim 1/(\Omega t)^{2\alpha{\epsilon^2\over\Omega^2}}$ for exponentially large times, where the power-law
exponent interpolates between $1$ for $\epsilon=0$ and $2\alpha$ for 
$\epsilon\gg\tilde{\Delta}$. The bias-dependent power-law exponent $2\alpha{\epsilon^2\over\Omega^2}$ has 
also been proposed in Ref.~\onlinecite{slutskin_etal_11} but we stress that it is only correct for
very large times and we will show that, for small times, other logarithmic contributions appear
which lead to a complicated crossover to a power-law
$\sim 1/(\Omega t)^{2\alpha}$ for exponentially small times. As already mentioned in 
Ref.~\onlinecite{kashuba_schoeller_13}, the determination of the correct long-time behavior of 
preexponential functions depends crucially on the vertex renormalization not taken into account in any
previous work. At zero bias this has lead to a correction of the NIBA-result \cite{kashuba_schoeller_13} and
we stress that all our results for nonzero bias presented in this paper can as well only be derived correctly by
including the vertex renormalization. 

The paper is organized as follows. In Section~\ref{sec:model_reference} we
introduce the ohmic spin boson model and the kinetic equation to calculate the time dynamics. 
We provide the perturbative expansion of the effective Liouvillian in Fourier space and explain 
its analytic structure together with the one of the reduced density matrix. We also provide the
propagator in renormalized perturbation theory which will be derived later in 
Section~\ref{sec:RTRG}. In Section~\ref{sec:time_dynamics} we will explicitly 
calculate the time dynamics in various time regimes. We review the exact solution at zero tunneling
and the Bloch-Redfield solution in Sections~\ref{sec:exact_solution} and \ref{sec:Bloch-Redfield}
as a reference. In Section~\ref{sec:perturbative} we present the results from renormalized perturbation 
theory and determine the time evolution in the regime of small times in Section~\ref{sec:small_times}
and in the whole regime where time is not exponentially small or large in 
Section~\ref{sec:non_exp_times}.
In Section~\ref{sec:RTRG} we use the RTRG method to account for all logarithmic terms nonperturbatively. 
In Section~\ref{sec:RG_equations} we set up the RG equations for the ohmic spin boson model and show 
in Sections~\ref{sec:RG_large_E} and \ref{sec:RG_non_exp_E} how the propagator has to be changed 
to account for all logarithmic renormalizations from high energies. The numerical solution of the
RG equations containing also all logarithmic renormalizations at low energies
will be presented in Section~\ref{sec:exp_large_times}. We close with a summary of our results 
in Section~\ref{sec:summary} and discuss their relevance for other models of dissipative quantum
mechanics. We use the unit $\hbar=1$ throughout this paper.

\section{Model, kinetic equation and Liouvillian}
\label{sec:model_reference}

In this Section we introduce the model under consideration and set up the kinetic 
equation to determine the time dynamics of the local reduced density
matrix. In addition we provide the perturbative solution for the effective Liouvillian
in Fourier space. This form is very helpful to understand the proper analytical continuation 
into the lower half of the complex plane and the correct procedure to avoid the occurrence of 
secular terms. Furthermore, we will present the perturbative determination of the decay poles.

\subsection{Model}
\label{sec:model}

The Hamiltonian for the spin boson model consists of a local $2$-level system
(described by Pauli matrices $\sigma_i$) coupled linearly to a bosonic bath with 
energy modes $\omega_q>0$
\begin{align}
\label{eq:H_tot}
H_{\text{tot}}\,&=\,H\,+\,H_{\text{bath}}\,+\,V\quad,\\
\label{eq:H}
H\,&=\,{\epsilon\over 2}\sigma_z\,-\,{\Delta\over 2}\sigma_x\quad,\\
\label{eq:H_res}
H_{\text{bath}}\,&=\,\sum_q \omega_q a_q^\dagger a_q\quad,\\
\label{eq:V}
V\,&=\,{1\over 2}\sigma_z\,\sum_q g_q(a_q+a_q^\dagger)\quad,
\end{align}
where $\epsilon$ denotes the bias, $\Delta$ the tunneling, and the coupling to the bath
is described by the coupling parameters $g_q$. We note that by a convenient spin rotation the 
coupling to the bath can always be chosen in the z-direction and the y-axis can be chosen 
perpendicular to the local spin in the Hamiltonian (the expectation value of the local spin
will of course get all components as function of time). The parameters $\epsilon$, $\Delta$ and
$g_q$ are real to guarantee hermiticity of $H_{\text{tot}}$ (please note that the sign convention for $\Delta$ 
is sometimes chosen differently in the literature). For convenience we choose $\Delta,\epsilon>0$ 
which again can always be achieved by an appropriate spin rotation.

The microscopic details of the modes $\omega_q$ and the coupling constants $g_q$ enter 
the time dynamics of the local system only via the energy dependence of the spectral density
\begin{align}
\label{eq:spectral_density}
J(\omega)\,=\,\pi\sum_q g_q^2\delta(\omega-\omega_q)\quad,
\end{align}
which, for the ohmic spin boson model, is parametrized as
\begin{align}
\label{eq:ohmic_spectral_density}
J(\omega)\,=\,2\pi\alpha\omega\theta(\omega)J_c(\omega)\quad,
\end{align}
where $\alpha$ is a dimensionless coupling constant and $J_c(\omega)$ is a high-energy 
cutoff function needed since frequency integrals diverge logarithmically at high energies
for all terms in the perturbative series in $\alpha$. In this paper we choose a
Lorentzian cutoff function (in contrast to exponential cutoffs $\sim e^{-\omega/D}$ often
used in the literature)
\begin{align}
\label{eq:cutoff_function}
J_c(\omega)\,=\,{D^2\over D^2 + \omega^2}\quad,
\end{align}
where $D$ denotes the band width. This choice is taken to simplify frequency integrals and 
influences only some prefactors of non-logarithmic terms but not the scaling behavior. The 
ohmic spin boson model in weak coupling is defined by the condition $\alpha\ll 1$ such that 
a perturbative expansion in $\alpha$ makes sense. 

Since we will also work in a basis where the local Hamiltonian $H$ is diagonal we introduce
the unitary transformation
\begin{align}
\label{eq:trafo}
U\,=\,U^\dagger\,=\,U^{-1}\,=\,{1\over\sqrt{2\Omega_0}}
\left(\begin{array}{cc} -v_- & v_+ \\ v_+ & v_- \end{array}\right)\quad,
\end{align}
where $v_\pm=\sqrt{\Omega_0\mp\epsilon}$ and 
\begin{align}
\label{eq:Omega_0}
\Omega_0\,=\,\sqrt{\epsilon^2 + \Delta^2}
\end{align}
denotes the bare level splitting (Rabi frequency) of the local system. With this unitary 
transformation we get $U H U^\dagger = {1\over 2}\Omega_0\sigma_z$, i.e. the eigenvalues 
$\pm\Omega_0/2$ with corresponding eigenvectors given by the two columns of $U^\dagger=U$. 

\subsection{Kinetic equation}
\label{sec:kinetic_equation}

We aim at calculating the time dynamics of the reduced density matrix of the local system
\begin{align}
\label{eq:rho(t)_definition}
\rho(t)\,=\,Tr_{\text{bath}}\rho_{\text{tot}}(t)
\end{align}
with an initial state for the total density matrix 
\begin{align}
\label{eq:rho_initial}
\rho_\text{tot}(t=0)\,=\,\rho_0\rho_\text{bath}^{\text{eq}}
\end{align}
factorizing into an arbitrary
initial state $\rho_0=\rho(t=0)$ for the local system and an equilibrium canonical distribution
$\rho_\text{bath}$ for the bath. For simplicity we set temperature $T=0$ in the following. 
Using standard projection operator \cite{projection_operator}, path integral \cite{weiss_12},
or diagrammatic \cite{schoeller_09,schoeller_14} techniques one can show that $\rho(t)$ can
be determined from a formally exact kinetic equation
\begin{align}
\label{eq:kinetic_time}
i\dot{\rho}(t)\,=\,L_0\rho(t) + \int_0^t dt' \Sigma(t-t')\rho(t')\quad,
\end{align}
where $L_0=[H,\cdot]$ and $\Sigma(t-t')$ are superoperators acting on operators. The first term on
the r.h.s. describes the time evolution from the von Neumann equation of the isolated local system,
whereas the second term contains the dissipative kernel $\Sigma(t-t')$ leading to irreversible time
dynamics into a stationary state $\rho_{\text{st}}=\lim_{t\rightarrow\infty}\rho(t)$. The various methods
described in Rfs.~\onlinecite{projection_operator,weiss_12,schoeller_09,schoeller_14} just differ in the
technique how to calculate this kernel in perturbation theory in $\alpha$. Since all quantities are
only defined for positive times, we define the Fourier transform as for retarded correlation functions
(for convenience we use the same symbol for the Fourier transform)
\begin{align}
\label{eq:fourier}
\rho(E)\,=\,\int_0^\infty dt \,e^{iEt} \rho(t)\quad,\quad
\Sigma(E)\,=\,\int_0^\infty dt \,e^{iEt} \Sigma(t)\quad,
\end{align}
which are well-defined analytic functions in the complex plane for all $E$ with positive 
imaginary part (a proper analytic continuation into the lower half of the complex plane will 
be discussed later). From (\ref{eq:kinetic_time}) we obtain the formal solution in Fourier space as
\begin{align}
\label{eq:solution_fourier}
\rho(E)\,=\,{i\over E-L(E)}\rho_0\quad,
\end{align}
where $L(E)=L_0+\Sigma(E)$ denotes the effective Liouvillian in Fourier space with matrix elements
$L_{s_1 s_2,s_1^\prime s_2^\prime}$ ($s$ denote the states of the local system). The Liouvillian has
the two important properties \cite{schoeller_09,schoeller_14}
\begin{align}
\label{eq:L_properties}
\text{Tr}L(E)\cdot\,=\,0\quad,\quad
L(E)^c\,=\,-L(-E^*)\quad,
\end{align}
where $\text{Tr}$ denotes the trace over the local system and the $c$-transform is defined by 
$L(E)^c_{s_1 s_2,s_1^\prime s_2^\prime}=L(E)_{s_2 s_1,s_2^\prime s_1^\prime}^*$.
From these properties one can show the conservation of probability $Tr\dot{\rho}(t)=0$ and the
hermiticity of the density matrix $\rho(t)^\dagger=\rho(t)$ \cite{schoeller_09,schoeller_14}.

Once $L(E)$ is known, the
time dynamics can be calculated from inverse Fourier transform as 
\begin{align}
\label{eq:solution_time}
\rho(t)\,=\,{i\over 2\pi}\int_{\cal{C}} dE {e^{-iEt}\over E-L(E)}\rho_0\quad,
\end{align}
where $\cal{C}$ is a straight line in the complex plane lying slightly above the real axis, i.e.
$E=x+i\eta$, with $\eta=0^+$ and $x$ running from $x=-\infty$ to $x=+\infty$ (the precise form of 
$\cal{C}$ in the upper half is not important since $\rho(E)$ is an analytic function there). We 
note that we have used the Fourier and not the Laplace transform (defined by $e^{-Et}$ in 
(\ref{eq:fourier})) since it makes the analogy to the analytic properties of retarded correlation
functions more transparent.    

As pointed out in detail in 
Rfs.~\onlinecite{pletyukhov_schuricht_schoeller_10,schoeller_09,schoeller_14,kashuba_schoeller_13} the most 
elegant way to determine the integral over $\cal{C}$ is to close the integration contour in the
lower half of the complex plane and to use a convenient analytic continuation of $L(E)$ and $\rho(E)$
into the lower half of the complex plane, such that all branch cuts point into the direction of the
negative imaginary axis and start at the branching points $z_i=\Omega_i-i\Gamma_i$. For the ohmic
spin boson model we note that $\rho(E)$ has one isolated pole at $E=0$ determining the stationary state from 
\begin{align}
\label{eq:stationary}
L(0^+)\rho_{\text{st}}\,=\,0\quad,
\end{align}
together with three branch cuts starting at the branching points (or poles)
\begin{align}
\label{eq:poles}
z_0\,=\,-i\Gamma\quad,\quad z_\pm=\pm\Omega-i\Gamma/2\quad,
\end{align}
with $\Gamma>0$, whereas $L(E)$ has only branch cuts starting at $z_0$ and $z_\pm$ without any poles. 
If we denote the eigenvalues of the $4\times 4$-matrix $L(E)$ by $\gamma_i(E)$ with $i=\text{st},0,\pm$, the
pole positions of the propagator $1/(E-L(E))$ follow from $\gamma_i(z_i)=z_i$ and it follows from 
(\ref{eq:L_properties}) that one eigenvalue must be zero and $-\gamma_i(E)^*$ are the eigenvalues of
$L(-E^*)$. Thus, $-\gamma_i(-E^*)^*$ must be also an eigenvalue of $L(E)$, leading to
\begin{align}
\label{eq:eigenvalue_st}
\gamma_{\text{st}}\,&=\,0\quad,\\
\label{eq:eigenvalue_0}
\gamma_0(E)\,&=\,-\gamma_0(-E^*)^*\quad,\\
\label{eq:eigenvalue_sigma}
\gamma_+(E)\,&=\,-\gamma_-(-E^*)^*\quad.
\end{align}
As a consequence, the pole $z_0$ is purely imaginary and $z_+=-z_-^*$, in accordance with (\ref{eq:poles}).

Using the diagrammatic technique
of Rfs.~\onlinecite{schoeller_09,schoeller_14,pletyukhov_schoeller_12,kashuba_schoeller_13} 
one can derive the analytic features in all orders of perturbation theory 
but it is illustrative to study them already from the perturbative solution for $L(E)$ up to $O(\alpha)$, which
will be presented in the next Section.

\subsection{Liouvillian in perturbation theory}
\label{sec:L_perturbative}

With the help of the diagrammatic technique used in Ref.~\onlinecite{kashuba_schoeller_13} for the
ohmic spin model at zero bias, we calculate the Liouvillian up to $O(\alpha)$ in 
Appendix~\ref{sec:appendix_L_perturbative}. Denoting the two states of the local system by $i=1,2$
(corresponding to the original Hamiltonian $H$ in (\ref{eq:H})) and using the sequence $(11,22,12,21)$ to 
numerate the matrix elements of superoperators, we find:
\begin{align}
\label{eq:L_sa}
L(E)\,&=\,L_0 + \Sigma_a(E) + \Sigma_s = L_a(E) + \Sigma_s\quad,\\
\label{eq:L_0}
L_0\,&=\,\left(\begin{array}{cc} 0 & \Delta\tau_- \\ \Delta\tau_- & \epsilon\sigma_z \end{array}\right)\quad,\\
\label{eq:sigma_s}
\Sigma_s\,&=\, i\pi\alpha\Delta 
\left(\begin{array}{cc} 0 & 0 \\ \tau_+ & 0  \end{array}\right)\quad,\\
\label{eq:sigma_a}
\Sigma_a(E)\,&=\,\alpha\sum_{i=0,\pm}{\cal{F}}_i(E)M_i\,\,,\,\,
M_i=\left(\begin{array}{cc} 0 & 0 \\ 0 & \hat{M}_i \end{array}\right)\quad,\\
\label{eq:hat_M_0}
\hat{M}_0\,&=\,2{\Delta^2\over\Omega_0^2}\tau_-\quad,\\
\label{eq:hat_M_sigma}
\hat{M}_\pm\,&=\,\tau_+\,\pm\,{\epsilon\over\Omega_0}\sigma_z \,+\, 
{\epsilon^2\over\Omega_0^2}\tau_-\quad,
\end{align}
where $\tau_\pm={1\over 2}(1\pm\sigma_x)$ and
\begin{align}
\label{eq:F}
{\cal{F}}_i(E)\,&=\,(E-\lambda_i(E)){\cal{L}}_i(E)\quad,\\
\label{eq:cal_L}
{\cal{L}}_i(E)\,&=\,\ln{-i(E-\lambda_i(E))\over D}\quad.
\end{align}
Here $\lambda_i(E)$ are the important functions
\begin{align}
\label{eq:lambda_0}
\lambda_0(E)\,&=\,-\alpha{\Delta^2\over\Omega_0}\sum_{\sigma=\pm}\sigma{\cal{L}}_\sigma(E)\quad,\\
\label{eq:lambda_sigma}
\lambda_\pm(E)\,&=\,\pm(\Omega_0+\alpha{\Delta^2\over\Omega_0}{\cal{L}}_0(E))\quad,
\end{align}
which determine the position of the poles (\ref{eq:poles}) of the resolvent $1/(E-L(E))$ 
(and also of $\rho(E)$ due to (\ref{eq:solution_fourier})) by solving the self-consistent equations
\begin{align}
\label{eq:determination_poles}
z_i\,=\,\lambda_i(z_i)\quad.
\end{align}
This can be seen from the derivation in Appendix~\ref{sec:appendix_L_perturbative}, where the
$\lambda_i(E)$ are defined as the eigenvalues of the Liouvillian $\tilde{L}_\Delta(E)$, defined by
\begin{align}
\label{eq:tilde_L}
\tilde{L}_\Delta(E)\,=\,Z'(E)L_\Delta(E)\quad,\quad Z'(E)\,=\,{1\over 1-L'(E)}\quad,
\end{align}
where $L_\Delta(E)$ and $L'(E)$ follow from the decomposition
\begin{align}
\nonumber
L_a(E)\,&=\,L_\Delta(E)\,+\,E L'(E)\\
\label{eq:L_decomposition}
\,&=\,L_0\,+\,\Sigma_\Delta(E)\,+\,E L'(E)\quad,
\end{align}
with
\begin{align}
\label{eq:sigma_delta}
\Sigma_\Delta(E)\,&=\,-\alpha\sum_{i=0,\pm}\lambda_i(E){\cal{L}}_i(E)M_i\quad,\\
\label{eq:L_prime}
L'(E)\,&=\,\alpha\sum_{i=0,\pm} {\cal{L}}_i(E)M_i\quad.
\end{align}
This decomposition is very
helpful since it exhibits the purely logarithmic superoperators $L_\Delta(E)$ and $L'(E)$, together with the 
terms linear in $E$. The eigenvalues of $L(E)$ and $\tilde{L}_\Delta(E)$ are different but the relation
(note that $\Sigma_s L_a=0$)
\begin{align}
\nonumber
{1\over E-L(E)}\,&=\,{1\over E-L_a(E)}(1+\Sigma_s{1\over E})\\
\label{eq:resolvent_tilde_L}
&=\,{1\over E-\tilde{L}_\Delta(E)}Z'(E)(1+\Sigma_s{1\over E})\quad,
\end{align}
shows that the poles of the two resolvents $1/(E-L(E))$ and $1/(E-\tilde{L}_\Delta(E))$ are the same, i.e.
the solutions $z_i$ of the self-consistent equations (\ref{eq:determination_poles}) provide indeed the 
nonzero poles of the resolvent $1/(E-L(E))$.

Most importantly, we see from the perturbative result (\ref{eq:L_sa})-(\ref{eq:sigma_a}) that $z_i$ 
are not only the poles of the local density matrix in Fourier space but at the same time determine the
branching points of the logarithmic functions ${\cal{L}}_i(E)$, i.e. determine the starting points for 
the branch cuts of $L(E)$ in the lower half of the complex plane. The logarithm in Eq.~(\ref{eq:cal_L}) is
the natural logarithm with a branch cut on the negative real axis, i.e. the branch cut w.r.t. the 
Fourier variable $E$ points into the direction of the negative imaginary axis, a choice which will be most
convenient for an analytical determination of the branch cut integral in the long time limit, see
Section~\ref{sec:time_dynamics}. The fact that the branching points of all logarithmic terms are
the same as the pole positions of the local density matrix is a very important observation and can be 
shown to hold in 
all orders of perturbation theory by using the diagrammatic method developed in
Rfs.~\onlinecite{pletyukhov_schuricht_schoeller_10,schoeller_09,schoeller_14,kashuba_schoeller_13}, see
also some remarks in Appendix~\ref{sec:appendix_L_perturbative}. 
Obviously, for this property it is very important to keep the functions $\lambda_i(E)$ in the argument of
the logarithm and not to expand ${\cal{L}}_i(E)$ in $\alpha$. As already mentioned in 
Ref.~\onlinecite{schoeller_14} in all detail, such an expansion leads to secular terms $(1/E)^n$ for 
the Liouvillian, e.g. for the expansion of $\alpha {\cal{F}}_0(E)$ one obtains
\begin{align}
\nonumber
\alpha {\cal{F}}_0(E)\,&=\,\alpha (E-\lambda_0(E))\ln{-iE\over D} - \alpha \lambda_0(E) \\
\label{eq:secular}
& + {1\over 2}\alpha \lambda_0(E)^2{1\over E} + O(\alpha^4)\quad.
\end{align}
We note that secular terms start at $O(\alpha^3)$ due to the factor $E-\lambda_0(E)$ in front of the
logarithm. Therefore, even in a calculation up to $O(\alpha^2)$ one can not see the occurrence of secular
terms in $L(E)$. 
The power of these secular terms increases with increasing order in $\alpha$ and, therefore, have to be
resummed nonperturbatively. They appear directly in the effective Liouvillian $L(E)$ and have to be 
distinguished from secular terms appearing by expanding the resolvent $1/(E-L_0-\Sigma(E))$ in $\Sigma(E)$.
The resummation of the latter are responsible to obtain the correct exponential behavior 
of the leading order Bloch-Redfield terms for the time evolution, whereas the ones in $L(E)$ have to be 
resummed to obtain the exponential part of all correction 
terms to the Bloch-Redfield solution. Essentially, the fact that logarithmic functions in all
orders of perturbation theory appear always in the form of ${\cal{L}}_i(E)$ is due to the property that
all bare propagators of the local system can be replaced by full propagators without any double counting,
see Appendix~\ref{sec:appendix_L_perturbative}. As a consequence the exact eigenvalues of $\tilde{L}_\Delta(E)$
appear in the perturbative series and {\it not} the bare ones. This fact is very important to notice in
order to find the correct nonanalytic features in the lower half of the complex plane. E.g. by calculating 
${\cal{F}}_0(E)$ only by the first term on the r.h.s. of (\ref{eq:secular}) one obtains a logarithm which has a 
branch cut starting at the origin leading to a term of the time evolution which is not exponentially 
decaying. The expansion (\ref{eq:secular}) is only well-defined for $E\sim\Omega_0$, i.e. on time scales
$t\sim 1/E \sim 1/\Omega_0$, where the solution is just oscillating and the decay has not yet set in.   
In this regime the perturbative solution of Ref.~\onlinecite{divincenzo_loss_05} can be used but 
{\it not} for larger time scales describing the crossover to the regime of exponential decay. 

We note that the perturbative solution (\ref{eq:L_sa})-(\ref{eq:sigma_a}) for $L(E)$ can only be used
when the logarithmic terms are small enough, i.e. the condition 
\begin{align}
\label{eq:E_condition_pert}
\alpha|\ln{-i(E-\lambda_i(E))\over D}|\,\ll\,1
\end{align}
should hold. This is obviously not fulfilled when $E$ approaches the branching point $z_i$ or is too far
away from it. Only the RG method presented in Section~\ref{sec:RTRG} is capable of resumming the logarithmic 
terms in all orders to find the correct scaling behavior for large $E$ or $E$ close to $z_i$. The condition 
(\ref{eq:E_condition_pert}) can be reformulated in terms of time by replacing $E-\lambda_i(E)\rightarrow 1/t$ 
leading to
\begin{align}
\label{eq:t_condition_pert}
\alpha|\ln(Dt)|\,\ll\,1\quad,
\end{align}
showing that the perturbative theory can not be used to calculate the time evolution for exponentially small or 
large times. However, as we will see in Section~\ref{sec:RTRG} these regimes can be studied as well by using the
RTRG method. 

As a consequence one should also not be concerned by the fact that the solution of the self-consistent
equations (\ref{eq:determination_poles}) with (\ref{eq:lambda_0}) and (\ref{eq:lambda_sigma}) is 
ill-defined due to the singularity of the logarithm. For times in the regime
(\ref{eq:t_condition_pert}) we need the functions $\lambda_i(E)$ only in the typical regime 
(\ref{eq:E_condition_pert}). Using $z_0\sim O(\alpha)$ and $z_\pm=\pm\Omega_0+O(\alpha)$, 
this means that for $|E-z_0|\sim\alpha^n\Omega_0$ (with some integer $n>0$) we can replace $\lambda_0(E)$ by
\begin{align}
\label{eq:lambda_0_close_to_z_0}
\lambda_0(E)\,\approx\,-\alpha{\Delta^2\over\Omega_0}\sum_{\sigma=\pm}\sigma\ln{-i(-\sigma\Omega_0)\over D}
\,=\,-i\Gamma_1\quad,
\end{align}
with
\begin{align}
\label{eq:gamma_1}
\Gamma_1\,=\,\pi\alpha{\Delta^2\over\Omega_0}\quad,
\end{align}
up to an error of $O(n\alpha^2\ln{\alpha})$. Up to the same error,
for $|E-z_\pm|\sim\alpha^n\Omega_0$, we can replace $\lambda_\pm(E)$ by
\begin{align}
\label{eq:lambda_sigma_close_to_z_sigma}
\lambda_\pm(E)\,\approx\,\pm(\Omega_0+\alpha{\Delta^2\over \Omega_0}\ln{-i(\pm\Omega_0)\over D})
\,=\,\pm\Omega_1-i\Gamma_1/2\quad,
\end{align}
with 
\begin{align}
\label{eq:omega_1}
\Omega_1\,=\,\Omega_0 - \alpha{\Delta^2\over\Omega_0}\ln{D\over\Omega_0}\quad.
\end{align}
Therefore, we conclude from the perturbative expansion that the solution of 
(\ref{eq:determination_poles}) is given by
\begin{align}
\label{eq:z_0}
z^{(1)}_0\,&=\,-i\Gamma_1 + O(\alpha^2\ln\alpha)\quad,\\
\label{eq:z_sigma}
z^{(1)}_\pm\,&=\,\pm\Omega_1-i\Gamma_1/2 + O(\alpha^2\ln\alpha)\quad.
\end{align}
In Section~\ref{sec:RTRG} we will resum all logarithmic renormalizations
$\sim(\alpha\ln(\Omega/D))^n$ from high energies and show that $\Omega_1$ has to be 
replaced by the renormalized Rabi frequency $\Omega$, which has the same form as $\Omega_0$
but the bare tunneling $\Delta$ has to be replaced by the renormalized tunneling 
$\tilde{\Delta}$ 
\begin{align}
\label{eq:omega_renormalized}
\Omega\,&=\,\sqrt{\epsilon^2 + \tilde{\Delta}^2}\quad,\\
\label{eq:tilde_delta}
\tilde{\Delta}\,&=\,\Delta\left({\Omega\over D}\right)^\alpha 
\,=\,\Delta\left({\sqrt{\epsilon^2+\tilde{\Delta}^2}\over D}\right)^\alpha \quad.
\end{align}
We note that the low-energy scale cutting off the logarithmic terms in this expression is set by
$\Omega$ but {\it not } by the renormalized tunneling as has been stated e.g. in Ref.~\onlinecite{weiss_12}. 
This was already mentioned in Ref.~\onlinecite{divincenzo_loss_05}, where the oscillation 
frequency has been calculated perturbatively up to the first logarithmic term, as given by 
Eq.~(\ref{eq:omega_1}). Furthermore, we note, that besides the logarithmic terms there can be 
other regular terms $\sim\alpha^n$ which depend on the specific high-energy cutoff function
under consideration. The logarithmic terms however are universal, i.e. do not depend on the
specific form of the high-energy cutoff function. This will be explained in Section~\ref{sec:RTRG}. 

Inserting the propagator (\ref{eq:resolvent_tilde_L}) in (\ref{eq:solution_time}) and using the
perturbative result (\ref{eq:L_sa}-\ref{eq:lambda_sigma}) for $L(E)$, one can systematically 
determine the time dynamics one order beyond Bloch-Redfield using the scheme presented in 
Section~\ref{sec:time_dynamics}. However, this calculation can be easily improved by using
renormalized perturbation theory, where the renormalized tunneling has to be used at appropriate
places and renormalizations of $Z$-factors are important. Therefore, although the 
explicit results will be derived later on in Section~\ref{sec:RTRG} using the RTRG method, 
we already state the results for the propagator in the next subsection such that we can use
it in Section~\ref{sec:time_dynamics} for the perturbative calculation of the time dynamics.

\subsection{Liouvillian in renormalized perturbation theory}
\label{sec:L_renormalized_perturbative}

In Section~\ref{sec:RTRG} we will show how the propagator $1/(E-L(E))$ has to be slightly modified to
account for all logarithmic renormalizations from high energies. There are two different kinds of 
logarithmic terms, one involving powers of $\alpha\ln(D/\Omega)$ (which can be resummed in the
renormalized tunneling (\ref{eq:tilde_delta})), the other containing powers of logarithmic terms 
$\alpha\ln(\Omega t)$ in time. The latter can be treated perturbatively provided that time is not
exponentially small or large. This defines the regime which we call the regime of \underline{times in 
the non-exponential regime} 
\begin{align}
\label{eq:non_exp_times}
|\alpha\ln(\Omega t)|\,\ll\,1\quad,
\end{align}
which corresponds in Fourier space to the regime
\begin{align}
\label{eq:non_exp_E}
|\alpha\ln{-i(E-z_i)\over \Omega}|\,\ll\,1\quad.
\end{align}
This is the regime where renormalized perturbation theory can be applied. 
In Section~\ref{sec:RG_non_exp_E} we will show that in this regime the propagator 
can be written as 
\begin{align}
\label{eq:renormalized_propagator}
{1\over E-L(E)}\,&\approx\,{1\over E-\tilde{L}_a(E)}Z^\prime(1+\Sigma_s{1\over E})\quad,
\end{align}
with
\begin{align}
\label{eq:Z'_non_exp_E}
Z^\prime\,&=\, \left(\begin{array}{cc} 1 & 0 \\ 0 & Z  \end{array}\right)\quad,\quad
Z\,=\,{\tilde{\Delta}^2\over\Delta^2}\\
\label{eq:tilde_L_a}
\tilde{L}_a(E)\,&=\,\tilde{L}_0 + \tilde{\Sigma}_a(E) \quad,\\
\label{eq:tilde_L_0}
\tilde{L}_0\,&=\,
\left(\begin{array}{cc} 0 & \Delta\tau_- \\ Z\Delta\tau_- & \epsilon\sigma_z \end{array}\right)\quad,\\
\label{eq:renormalized_sigma_a}
\tilde{\Sigma}_a(E)\,&=\,\alpha\sum_{i=0,\pm}{\cal{F}}_i(E)M_i\,\,,\,\,
M_i=\left(\begin{array}{cc} 0 & 0 \\ 0 & \hat{M}_i \end{array}\right)\quad,\\
\label{eq:renormalized_hat_M_0}
\hat{M}_0\,&=\,2{\tilde{\Delta}^2\over\Omega^2}\tau_-\quad,\\
\label{eq:renormalized_hat_M_sigma}
\hat{M}_\pm\,&=\,\tau_+\,\pm\,{\epsilon\over\Omega}\sigma_z \,+\, 
{\epsilon^2\over\Omega^2}\tau_-\quad,
\end{align}
where ${\cal{F}}_i(E)$ is defined by (\ref{eq:F}) with
\begin{align}
\label{eq:renormalized_lambda_0}
\lambda_0(E)\,&=\,-\alpha{\tilde{\Delta}^2\over\Omega}\sum_{\sigma=\pm}\sigma{\cal{L}}_\sigma(E)\quad,\\
\label{eq:renormalized_lambda_sigma}
\lambda_\pm(E)\,&=\,\pm(\Omega+\alpha{\tilde{\Delta}^2\over\Omega}{\cal{L}}_0(E))\quad,
\end{align}
and 
\begin{align}
\label{eq:cal_L_Omega}
{\cal{L}}_i(E)\,&=\,\ln{-i(E-\lambda_i(E))\over \Omega}\quad.
\end{align}
In comparison to the unrenormalized perturbation theory (\ref{eq:L_sa}-\ref{eq:lambda_sigma}) we see that the
renormalized Rabi frequency and the renormalized tunneling appear in $\tilde{\Sigma}_a(E)$ and $\lambda_i(E)$
instead of the bare ones and the band width $D$ is replaced by $\Omega$ in the logarithmic function 
${\cal{L}}_i(E)$. In addition, $L_0$ and the propagator
get a renormalization from the $Z'$-matrix containing the $Z$-factor $Z=\tilde{\Delta}^2/\Delta^2$. 
In Section~\ref{sec:RTRG} we will see that $Z$ can be obtained from a poor man scaling equation 
for $Z(E)=(-iE/D)^{2\alpha}$ cut off at $E=i\Omega$. Our result 
shows that renormalized perturbation theory is {\it not} obtained by just replacing 
$\Delta\rightarrow\tilde{\Delta}$ defining a local system with a renormalized tunneling. Instead, the
Liouvillian $\tilde{L}_0$ is no longer hermitian, i.e. can essentially be {\it not} expressed as a 
commutator with a renormalized local Hamiltonian. 

Since the solutions of $z_i=\lambda_i(z)$ again define the positions of the poles of the propagator, 
the logarithmic renormalizations from high energies lead in analogy to (\ref{eq:z_0}) and (\ref{eq:z_sigma})
to the renormalized pole positions
\begin{align}
\label{eq:z_0_RG}
z_0\,&=\,-i\Gamma + O(\alpha^2)\quad,\\
\label{eq:z_sigma_RG}
z_\pm\,&=\,\pm\Omega-i\Gamma/2 + O(\alpha^2)\quad,
\end{align}
with 
\begin{align}
\label{eq:gamma_RG}
\Gamma\,=\,\pi\alpha{\tilde{\Delta}^2\over\Omega}\quad.
\end{align}

In Section~\ref{sec:RTRG} we will also discuss the regimes of \underline{exponentially small or large
times} where the condition (\ref{eq:non_exp_times}) fails and higher powers of logarithmic 
terms have to be resummed by a proper RG method for the ultraviolett regime (small times or
large energies) and the infrared regime (large times or energies close to the pole positions). 
Although this regime is certainly of minor interest to quantum information processing it is
of high interest from a theoretical point of view since various power-laws appear which are
qualitatively very different in the ultraviolett and infrared regime. Furthermore, these
power laws are not only of academic interest in unrealistic time regimes since they become 
clearly visible for moderate $\alpha\sim 0.05$ and, moreover, second
order terms $\sim(\alpha\ln(\Omega t))^2$ can become of order $\alpha$ already for time
scales $t\sim 1/\Gamma$ where the decay is still moderate depending on the ratio of 
$\Omega/\tilde{\Delta}$. Using (\ref{eq:gamma_RG}) we find for $t\sim 1/\Gamma$
\begin{align}
\label{eq:time_scales_exp}
(\alpha\ln(\Omega t))^2\sim \alpha \quad\Leftrightarrow\quad
{\Omega\over\tilde{\Delta}}\sim\sqrt{\pi\alpha}\,\,e^{1/(2\sqrt{\alpha})}\quad,
\end{align}
leading e.g. to $\Omega/\tilde{\Delta}\sim 4$ for $\alpha=0.05$. This are quite realistic values
showing that higher powers of logarithmic terms contribute significantly on the same level as
corrections $\sim\alpha$ to the Bloch-Redfield solution. Although the terms $\sim\alpha\ln(\Omega t)$
are the leading order terms in this regime, the second order terms $\sim(\alpha\ln(\Omega t))^2$ are
clearly visible in the time dynamics of the preexponential functions showing a significant 
deviation from a straight line plotted logarithmically as function of $\ln(\Omega t)$, see 
Section~\ref{sec:exp_large_times}. Thus, for the spin boson model at finite
bias, the systematic calculation of corrections to Bloch-Redfield is quite subtle and requires
an analysis of higher-order terms beyond $O(\alpha)$ for the Liouvillian for various reasons.

In Section~\ref{sec:RTRG} we will see that the resummation of logarithmic terms in time is very
complicated in the infrared regime and requires a careful solution of the full RG equations, which
we will perform numerically. In contrast, the resummation of time-dependent logarithmic terms in the 
ultraviolett regime is quite straightforward since, for large energies, the energy scales of the
local system do not play an important role and can be treated perturbatively. Therefore, we state
here also the result for the propagator in the regime of {\underline{small times}} defined by
\begin{align}
\label{eq:small_times}
{1\over D}\,\ll\,t\,\ll\,{1\over \Omega}\quad,
\end{align}
corresponding to the regime of large energies 
\begin{align}
\label{eq:large_E}
\Omega \,\ll\, |E| \,\ll\, D \quad.
\end{align}
We note that resumming all logarithmic terms $\sim(\alpha\ln(E/D))^n$ or $\sim(\alpha\ln(Dt))^n$ leads to a 
universal result for the time evolution in the regime $|\alpha\ln(\Omega t)|\sim 1$ and $t\gg 1/D$ 
(where all corrections of $O(\alpha)$ and $O(1/(Dt)$ can be neglected), in contrast to the 
non-universal regime $t\lesssim 1/D$, where
bare perturbation theory in $\alpha$ can be used to determine $\rho(t)$ and the result depends crucially
on the shape of the high-energy cutoff function $J_c(\omega)$.   

For large energies $E\sim 1/t \gg \Omega$, we neglect all terms 
of relative order $\alpha\Omega/E\sim \alpha\Omega t$ in $\tilde{L}_\Delta(E)$
and $Z'(E)$ and find in Section~\ref{sec:RG_large_E} that
\begin{align}
\label{eq:tilde_L_delta_large_energies}
\tilde{L}_\Delta(E)\,&\approx\,\tilde{L}_0(E)\left(1\,+\,O(\alpha\Omega/E)\right)\quad,\\
\label{eq:Z'_large_E}
Z^\prime(E)\,&\approx\, \left(\begin{array}{cc} 1 & 0 \\ 
0 & Z(E)  \end{array}\right)
\left(1\,+\,O(\alpha\Omega/E)\right)\quad,
\end{align}
with
\begin{align}
\label{eq:tilde_L_0_large_energies}
\tilde{L}_0(E)\,&=\,\left(\begin{array}{cc} 0 & \Delta\tau_- \\ 
\Delta Z(E)\tau_- & \epsilon\sigma_z \end{array}\right)\quad,\\
\label{eq:Z_large_E}
Z(E)\,&=\,\left({-iE\over D}\right)^{2\alpha}\quad.
\end{align}
Since $\Sigma_s/E\sim\alpha\Omega/E$ can also be neglected in
(\ref{eq:resolvent_tilde_L}) we find for the propagator the approximation
\begin{align}
\label{eq:propagator_high_E}
{1\over E-L(E)}\,\approx\,{1\over E - \tilde{L}_0(E)} 
\left(\begin{array}{cc} 1 & 0 \\ 0 & Z(E)  \end{array}\right)\quad.
\end{align}
In Section~\ref{sec:RG_large_E} we will see that the form for $Z'(E)$ results from a poor man scaling
equation cut off at the largest energy scale $E$ which corresponds to $1/t$ in time space. If $E$ 
becomes of the order $\Omega$ the $Z$-factor is cut off at $E=i\Omega$, leading to the $Z$-factor
(\ref{eq:Z'_non_exp_E}) used in the regime where time is not exponentially small or large.   

The form (\ref{eq:propagator_high_E}) can be used in the whole regime $\Omega\ll E\ll D$, irrespective
of whether $E$ is exponentially large or not. Thus, we can also use it in the regime where
$|\alpha\ln(-iE/\Omega)|\ll 1$, where we can expand $Z(E)$ as
\begin{align}
\label{eq:perturbative_expansion_Z}
Z(E)\,=\,{\tilde{\Delta}^2\over\Delta^2}\,\left(1+2\alpha\ln{-iE\over\Omega}\right)\quad,
\end{align}
and, after a straightforward calculation, one finds that the propagator (\ref{eq:propagator_high_E})
at high energies obtains the same form in leading order in $\alpha$ and $\Omega/E$ as the propagator 
(\ref{eq:renormalized_propagator}) in the regime of non-exponentially large energies.

\section{Time dynamics}
\label{sec:time_dynamics}

In this section we will present the time dynamics of the local density matrix analytically in the
regimes of small times (including the case of exponentially small times) and for the regime of
times which are not exponentially small or large, where renormalized perturbation theory can be applied using
the propagator presented in Section~\ref{sec:L_renormalized_perturbative}. The exact solution
for zero tunneling and the lowest order Bloch-Redfield solution will be rederived 
in Sections~\ref{sec:exact_solution} and \ref{sec:Bloch-Redfield} for reference. 
In Section~\ref{sec:perturbative} we will present renormalized perturbation theory 
to show how the Bloch-Redfield solution has to be modified, together with the systematic calculation of the 
next correction in $O(\alpha)$. For the most interesting regime of times which are not exponentially 
small or large we note that our analytic solution has never been obtained correctly 
in the literature before.

\subsection{Exact solution at zero tunneling}
\label{sec:exact_solution}

For zero tunneling the time dynamics can be calculated exactly even for an arbitrary 
spectral density and finite temperatures \cite{leggett_87,weiss_12}. In this case the
local Hamiltonian $H=\sigma_z\epsilon/2$ decouples from the rest and the coupling to the bath 
can be eliminated by a unitary transformation shifting the field operators of the bath
\begin{align}
\label{eq:H_tranformed}
H_{\text{tot}}\,&=\,H + e^{\sigma_z \chi}H_{\text{res}}e^{-\sigma_z\chi} + c\quad,\\
\label{eq:shift_operator}
\chi\,&=\,\sum_q g_q (a_q + a_q^\dagger)\quad,
\end{align}
with an unimportant constant $c=\sum_q\omega_q g_q^2$ dropping out for the time dynamics. After a 
straightforward calculation, the time dynamics for the diagonal and non-diagonal matrix elements of
$\rho(t)$ follows as
\begin{align}
\label{eq:rho_diagonal_zero_tunneling}
\rho(t)_{\sigma\sigma}\,&=\,\rho(0)_{\sigma\sigma}\quad,\\
\label{eq:rho_nondiagonal_zero_tunneling}
\rho(t)_{\sigma,-\sigma}\,&=\,e^{-i\sigma\epsilon t}\langle e^{2(\chi(t)-\chi)}\rangle_{\text{res}}
\rho(0)_{\sigma,-\sigma}\quad,
\end{align}
where $\sigma=\pm\equiv 1,2$ denotes the two local states, 
$\chi(t)$ is the Heisenberg picture w.r.t. $H_{\text{res}}$, and $\langle\cdots\rangle_{\text{res}}$
denotes the expectation value w.r.t. to the canonical equilibrium distribution of the reservoir. 
Calculating this average by standard means gives the following result for the expectation values of
the Pauli matrices of the local system
\begin{align}
\label{eq:sigma_x_zero_tunneling}
\langle\sigma_x\rangle(t)\,&=\,e^{-h(t)}\,\left\{\cos(\epsilon t)\langle\sigma_x\rangle(0)
-\sin(\epsilon t)\langle\sigma_y\rangle(0)\right\}\quad,\\
\label{eq:sigma_y_zero_tunneling}
\langle\sigma_y\rangle(t)\,&=\,e^{-h(t)}\,\left\{\sin(\epsilon t)\langle\sigma_x\rangle(0)
+\cos(\epsilon t)\langle\sigma_y\rangle(0)
\right\}\quad,\\
\label{eq:sigma_z_zero_tunneling}
\langle\sigma_z\rangle(t)\,&=\,\langle\sigma_z\rangle(0)\quad,
\end{align}
with
\begin{align}
\label{eq:h(t)}
h(t)\,=\,{1\over\pi}\int_0^\infty d\omega J(\omega)(1+2 n(\omega)){1-\cos(\omega t)\over \omega^2}\quad,
\end{align}
where $n(\omega)$ is the Bose distribution function which vanishes at zero temperature. Thus, at zero
temperature, we get for the ohmic case 
\begin{align}
\label{eq:zero_tunneling_ohmic}
h(t)\,=\,2\alpha \int_0^\infty d\omega J_c(\omega) {1-\cos(\omega t)\over\omega}\quad,
\end{align}
which contains a logarithmic divergence at large $\omega$. Therefore, in the limit $Dt\gg 1$, we get
the result
\begin{align}
\label{eq:zero_tunneling_ohmic_large_t}
h(t)\,\approx\,2\alpha(\gamma + \ln(Dt))\quad,
\end{align}
where $\gamma$ is Euler's constant. This leads to the universal power-law 
\begin{align}
\nonumber
e^{-h(t)}\,&\approx\,(1-2\alpha\gamma)\left({1\over Dt}\right)^{2\alpha}\\
\label{eq:zero_tunneling_power_law}
&=\,(1-2\alpha\gamma){\tilde{\Delta}^2\over\Delta^2}\left({1\over \Omega t}\right)^{2\alpha}
\end{align}
for the time dynamics, where we have written the factor in front up to $O(\alpha)$ in order to compare
it later on to our perturbative solution for arbitrary tunneling. For the second form we have used
(\ref{eq:tilde_delta}) to write the result independent of $D$ parametrizing it by the ratio of the renormalized 
tunneling to the unrenormalized one (which is finite even in the limit of zero tunneling). 

As one can see the result (\ref{eq:zero_tunneling_power_law}) contains a resummation of all powers of 
logarithmic terms $\sim(\alpha\ln(Dt))^n$ and, thus, can only be obtained from the RG procedure presented in
Section~\ref{sec:RTRG}. It will turn out that it holds even at finite tunneling $\tilde{\Delta}\ll\epsilon$, 
provided that the condition $\Omega t\gg 1 \gg \Gamma t$ holds.

\subsection{Bloch-Redfield solution}
\label{sec:Bloch-Redfield}
The easiest way to derive the Bloch-Redfield solution is to insert
(\ref{eq:resolvent_tilde_L}) in (\ref{eq:solution_time}) and use the spectral 
decomposition of the Liouvillian $\tilde{L}_\Delta(E)$. This gives the formally exact expression
\begin{align}
\nonumber
\rho(t)\,&=\,{i\over 2\pi}\sum_{i=\text{st},0,\pm}\int_{\cal{C}} dE {e^{-iEt}\over E-\lambda_i(E)}\cdot\\
\label{eq:starting_point_poles}
&\hspace{2cm}
\cdot P_i(E)Z'(E)(1+\Sigma_s{1\over E})\rho_0\quad.
\end{align}
Here, $\lambda_i(E)$ are the eigenvalues of $\tilde{L}_\Delta(E)$ and $P_i(E)$ are the 
corresponding projectors. These quantities can be calculated by solving for the right and left
eigenstates of $\tilde{L}_\Delta(E)$
\begin{align}
\label{eq:right_eigenstate}
\tilde{L}_\Delta(E)|x_i(E)\rangle\,&=\,\lambda_i(E)|x_i(E)\rangle\quad,\\
\label{eq:left_eigenstate}
\langle\bar{x}_i(E)|\tilde{L}_\Delta(E)\,&=\,\langle\bar{x}_i(E)|\lambda_i(E)\quad,\\
\label{eq:projector}
P_i(E)\,&=\,|x_i(E)\rangle\langle \bar{x}_i(E)|\quad.
\end{align}
The projectors fulfill the property
\begin{align}
\label{eq:eigenvectors_property}
P_i(E) P_j(E)\,=\,\delta_{ij}P_i(E)\quad,\quad
\sum_i P_i(E)\,=\,1\quad.
\end{align}
We note that the eigenvalues are complex since the superoperator $\tilde{L}_\Delta(E)$ is a
non-hermitian matrix. One of the eigenvalues is zero (denoted by $i=\text{st}$) 
and the corresponding right/left eigenstates are exactly known in all orders of perturbation 
theory 
\begin{align}
\label{eq:st_eigenvalue}
\lambda_{\text{st}}\,&=\,0 \quad,\\
\label{eq:st_eigenstates}
|x_{\text{st}}\rangle\,&=\,{1\over\sqrt{2}}
\left(\begin{array}{c} 1 \\ 1 \\ 0 \\ 0 \end{array}\right)\quad,\quad
\langle \bar{x}_{\text{st}}|\,=\,{1\over\sqrt{2}}
\left(\begin{array}{cccc} 1 & 1 & 0 & 0 \end{array}\right)\\
\label{eq:P_st}
P_{\text{st}}\,&=\,
\left(\begin{array}{cc} \tau_+ & 0 \\ 0 & 0 \end{array}\right)\quad.
\end{align}
This can be seen from the matrix structure (\ref{eq:sigma_delta}-\ref{eq:L_prime}), which holds in all
orders of perturbation theory, see Appendix~\ref{sec:appendix_L_perturbative} for the proof. 
We note that the right eigenstate $|x_{\text{st}}(E)\rangle$ for $E=0^+$ does not give the stationary state 
$\rho_{\text{st}}$, following from (\ref{eq:stationary}), since the eigenstates of 
$\tilde{L}_\Delta(E)$ and $L(E)$ are different.

The eigenvalues $\lambda_i(E)$ for $i=0,\pm$ have already been provided in
perturbation theory up to $O(\alpha)$ in (\ref{eq:lambda_0}) and (\ref{eq:lambda_sigma}). 
Since $P_{\text{st}}Z'(E)=P_{\text{st}}$ and $\langle\bar{x}_{\text{st}}|\Sigma_s=0$ we note that
the second term involving $\Sigma_s$ contributes only for $i\ne\text{st}$. 

The Bloch-Redfield solution is obtained by taking $P_i(E)Z'(E)\approx P_i^{(0)}$ in lowest order in 
$\alpha$ (which is independent of $E$) and taking the Markovian approximation
$\lambda_i(E)\approx\lambda_i(z_i)=z_i$, which again neglects $O(\alpha)$ contributions from the
residua and further corrections arising from possible branch cuts starting at $z_i$. The pole positions
are taken from (\ref{eq:z_0}), (\ref{eq:z_sigma}) and $z_{\text{st}}=0$. This gives the result
\begin{align}
\nonumber
\rho^{(0)}(t)\,&=\,{i\over 2\pi}\sum_{i=\text{st},0,\pm}\int_{\cal{C}} dE {e^{-iEt}\over E-z_i^{(1)}}
P_i^{(0)}(1+\Sigma_s{1\over E})\rho_0\\
\nonumber
\,&=\,(e^{-iz^{(1)}t}-1){1\over z^{(1)}_0} P_0^{(0)} \Sigma_S \rho_0\,+\,P_{\text{st}}\rho_0\\
\label{eq:Bloch_Redfield}
&\hspace{1cm}
\,+\,\sum_{i=0,\pm} e^{-iz^{(1)}_it} P_i^{(0)}\rho_0\quad.
\end{align}
The first term on the r.h.s. arises from the pole at $E=0$ 
from the term $\Sigma_s/E$. It is of 
$O(1)$ since $1/z^{(1)}_0=i/\Gamma_1\sim 1/\alpha$, in contrast
to the contributions from $1/z^{(1)}_\pm=1/(\pm\Omega_1-i\Gamma_1/2)\sim O(1)$ 
which lead to an $O(\alpha)$ correction to $\rho(t)$.

The projectors in lowest order are the ones for $L_0$. We note that there is no problem with
degenerate perturbation theory for the two eigenvalues $\lambda_{\text{st}}=0$ and $\lambda_0\sim\alpha$
(requiring in general a knowledge of $\tilde{L}_\Delta$ up to $O(\alpha)$ to calculate $P_{\text{st}}$ 
and $P_0$ in lowest order) since the projector $P_\text{st}$ is exactly known from (\ref{eq:P_st}) 
in all orders of perturbation theory such that $P_0^{(0)}=1-P_{\text{st}}-P_+^{(0)}-P_-^{(0)}$ can be used. 
The projectors for $L_0$ can be most easily obtained by transforming the matrix $L_0$ to the basis
of the exact eigenstates of $H$, which, by using the unitary matrix (\ref{eq:trafo}), is described 
by the unitary transformation $(A_0)_{ij,kl}=U_{ik}U_{jl}^*$ leading to
\begin{align}
\label{eq:matrix_A_0}
A_0\,=\,A_0^\dagger\,=\,A_0^{-1}\,=\,
{1\over \Omega_0}
\left(\begin{array}{cc} \Omega_0\tau_++\epsilon\tau_- & -\Delta\sigma_z\tau_+ \\ 
-\Delta\sigma_z\tau_- & -\epsilon\tau_+-\Omega_0\tau_- \end{array}\right).
\end{align}
In the new basis $L_0$ is given by
\begin{align}
\label{eq:L_0_transformed}
A_0 L_0 A_0^\dagger\,=\,
\left(\begin{array}{cc} 0 & 0 \\
0 & \Omega_0\sigma_z \end{array}\right)\quad,
\end{align}
and the projectors obviously follow from
\begin{align}
\label{eq:P_st_tilde}
A_0 P_{\text{st}}A_0^\dagger\,&=\,
\left(\begin{array}{cc} \tau_+ & 0 \\ 0 & 0 \end{array}\right)\quad,\\
\label{eq:P_0_tilde_lowest_order}
A_0 P_{0}^{(0)} A_0^\dagger \,&=\,
\left(\begin{array}{cc} \tau_- & 0 \\ 0 & 0 \end{array}\right)\quad,\\
\label{eq:P_sigma_tilde_lowest_order}
A_0 P_{\sigma}^{(0)} A_0^\dagger \,&=\,
\left(\begin{array}{cc} 0 & 0 \\ 0 & {1\over 2}(1+\sigma\sigma_z) \end{array}\right)\quad.
\end{align}
Transforming back with the matrix $A_0$ to the original basis one obtains straightforwardly
the result 
\begin{align}
\label{eq:P_0_lowest_order}
P_{0}^{(0)}\,&=\,{1\over \Omega_0^2}
\left(\begin{array}{cc} \epsilon^2\tau_- & -\Delta\epsilon\sigma_z\tau_+ \\ 
-\Delta\epsilon\sigma_z\tau_- & \Delta^2\tau_+ \end{array}\right)\quad,\\
\nonumber
P_{\sigma}^{(0)}\,&=\,{1\over 2\Omega_0^2}
\left(\begin{array}{cc} \Delta^2\tau_- & \Delta\epsilon\sigma_z\tau_+ \\ 
\Delta\epsilon\sigma_z\tau_- & \epsilon^2\tau_+ + \Omega_0^2\tau_- \end{array}\right)\,+\\
\label{eq:P_sigma_lowest_order}
&\hspace{1cm}+\,{\sigma\over 2\Omega_0}
\left(\begin{array}{cc} 0 & \Delta\tau_- \\ \Delta\tau_- & \epsilon\sigma_z \end{array}\right)\quad.
\end{align}

Inserting (\ref{eq:P_st}), (\ref{eq:P_0_lowest_order}), (\ref{eq:P_sigma_lowest_order}) and
(\ref{eq:sigma_s}) in (\ref{eq:Bloch_Redfield}) we obtain the Bloch-Redfield solution. Using the
formulas (\ref{eq:z_0}) and (\ref{eq:z_sigma}) for the pole positions, we can 
decompose the time evolution of the Pauli matrices generically as
\begin{align}
\nonumber
\langle\sigma_\alpha\rangle(t)\,&=\,
\langle\sigma_\alpha\rangle_{\text{st}}\,+\,
F^0_\alpha(t)e^{-\Gamma_1 t}\,+\,\\
\label{eq:time_dynamics_pauli_general_form}
&+\,F^c_\alpha(t)e^{-{\Gamma_1\over 2} t}\cos(\Omega_1 t)\,+\,
F^s_\alpha(t)e^{-{\Gamma_1\over 2} t}\sin(\Omega_1 t)\quad,
\end{align}
with $\alpha=x,y,z$. $F_\alpha^{0,c,s}(t)$ denote the preexponential functions, which 
become time independent in Bloch-Redfield approximation 
\begin{align}
\label{eq:sigma_st}
{\langle\sigma_x\rangle}_{\text{st}}\,&=\,{\Delta\over\Omega_0}\quad,\quad
{\langle\sigma_y\rangle}_{\text{st}}\,=\,0 \quad,\quad  
{\langle\sigma_z\rangle}_{\text{st}}\,=\,-{\epsilon\over\Omega_0} \quad,\\
\label{eq:F_0_x}
F^0_x\,&=\,-{\langle\sigma_x\rangle}_{\text{st}}-{\Delta\over\Omega_0}
{\langle\sigma_z^\prime\rangle}_0\quad,\quad\\
\label{eq:F_0_y}
F^0_y\,&=\,0 \quad,\quad\\
\label{eq:F_0_z}
F^0_z\,&=\,-{\langle\sigma_z\rangle}_{\text{st}}+{\epsilon\over\Omega_0}
{\langle\sigma_z^\prime\rangle}_0\quad,\\
\label{eq:F_c_x}
F^c_x\,&=\,-{\epsilon\over\Omega_0}{\langle\sigma_x^\prime\rangle}_0\quad,\\
\label{eq:F_c_y}
F^c_y\,&=\,{\langle\sigma_y\rangle}_0 \quad,\\
\label{eq:F_c_z}
F^c_z\,&=\,-{\Delta\over\Omega_0}{\langle\sigma_x^\prime\rangle}_0\quad,\\
\label{eq:F_s_x}
F^s_x\,&=\,-{\epsilon\over\Omega_0}{\langle\sigma_y\rangle}_0\quad,\quad\\
\label{eq:F_s_y}
F^s_y\,&=\,-{\langle\sigma_x^\prime\rangle}_0\quad,\\
\label{eq:F_s_z}
F^s_z\,&=\,-{\Delta\over\Omega_0}{\langle\sigma_y\rangle}_0\quad,
\end{align}
where
\begin{align}
\label{eq:sigma_x_prime}
\sigma_x^\prime\,&=\,
-{1\over\Omega_0}\Big(\epsilon\sigma_x + \Delta\sigma_z\Big)\quad,\\
\label{eq:sigma_y_prime}
\sigma_y^\prime\,&=\,-\sigma_y\quad,\\
\label{eq:sigma_z_prime}
\sigma_z^\prime\,&=\,{1\over\Omega_0}\Big(\epsilon\sigma_z - \Delta\sigma_x\Big)
\end{align}
are the Pauli spin operators in the basis where the local Hamiltonian is diagonal.

For later reference, we also state the form of the Bloch-Redfield solution
in the regime of small times where $\Omega_0 t\ll 1$. Expanding the exponentials up to linear
order in $\Omega_1 t$ and neglecting $\Gamma_1t,(\Omega_1-\Omega_0)t\sim \alpha\Delta t$  
we obtain
\begin{align}
\label{eq:sigma_x_small_times_Bloch_Redfield}
\langle\sigma_x\rangle(t)\,&=\,
{\langle\sigma_x\rangle}_0 - \epsilon t{\langle\sigma_y\rangle}_0\quad,\\
\label{eq:sigma_y_small_times_Bloch_Redfield}
\langle\sigma_y\rangle(t)\,&=\,
{\langle\sigma_y\rangle}_0 + \epsilon t{\langle\sigma_x\rangle}_0
+ \Delta t{\langle\sigma_z\rangle}_0\quad,\\
\label{eq:sigma_z_small_times_Bloch_Redfield}
\langle\sigma_z\rangle(t)\,&=\,{\langle\sigma_z\rangle}_0 
- \Delta t{\langle\sigma_y\rangle}_0\quad.
\end{align}

\subsection{Renormalized perturbation theory}
\label{sec:perturbative}

Using the propagators provided in Section~\ref{sec:L_renormalized_perturbative} we will now apply
renormalized perturbation theory to calculate the modification of the Bloch-Redfield solution in
lowest order in $\alpha$ (but including all logarithmic corrections $\sim(\alpha\ln(Dt))^n$ and
$\sim(\alpha\ln(D/\Omega))^n$ from high energies in all orders) together with the first systematic
correction in $O(\alpha)$ to the Bloch-Redfield solution. Since renormalized perturbation theory
can only be applied analytically in the regimes of small times or times which are not exponentially
small or large, we will restrict our analysis to these two regimes and find that the two solutions
coincide for small but not exponentially small times, so that also the crossover between these two
regimes can be described with our analytic results, providing a systematic analytic solution beyond
Bloch-Redfield in the most interesting regime for quantum information where the exponential decay 
has not yet destroyed the time dynamics completely. Only the regime of very large times where higher 
powers of logarithmic terms like $\alpha^2\ln^2(\Omega t)$ become important is not treated analytically 
and will be presented in Section~\ref{sec:exp_large_times} via a numerical solution of the RG equations.

\subsubsection{Small times}
\label{sec:small_times}

For small times $\Omega t \ll 1$ but still in the universal regime $t\gg 1/D$ we 
take the form (\ref{eq:propagator_high_E}) for the propagator and, since
$E\sim 1/t\gg \epsilon,\Delta$, can expand the resolvent up to first order in $\tilde{L}_0(E)$
\begin{align}
\label{eq:expansion_propagator_small_times}
{1\over E - \tilde{L}_0(E)} \,\approx\,
{1\over E}+{1\over E}\tilde{L}_0(E){1\over E}\quad.
\end{align}
In this way we keep all terms $\sim \tilde{L}_0/E\sim \epsilon t,\Delta t$, which, for $\Omega t\sim \alpha$,
can be of the same order as the first correction $\sim\alpha$ to the Bloch-Redfield result.

Inserting (\ref{eq:expansion_propagator_small_times}) and (\ref{eq:propagator_high_E}) in
(\ref{eq:solution_time}), using the integrals
\begin{align}
\nonumber
I_1(t)\,&=\,{i\over 2\pi}\int_{\cal{C}} dE e^{-iEt}{Z(E)\over E}\\
\label{eq:integral_I_1}
&=\, {\sin(2\pi\alpha)\over 2\pi\alpha}\Gamma(1+2\alpha)\left({1\over Dt}\right)^{2\alpha}\\
\nonumber
I_2(t)\,&=\,{i\over 2\pi}\int_{\cal{C}} dE e^{-iEt}{Z(E)\over E^2}\\
\label{eq:integral_I_2}
&=\, {-it\over 1-2\alpha}I_1(t)\quad,
\end{align}
where $\Gamma(x)$ denotes the Gamma function with 
$\Gamma(1+x)=1-\gamma x+O(x^2)$, and neglecting all terms of $O(\alpha^2)$, 
$O(\alpha\epsilon t)$ and $O(\alpha\Delta t)$, we find for the local density matrix the 
following result in the short time limit
\begin{align}
\nonumber
\rho(t)\,&=\,\left(\begin{array}{cc} 1 & 0 \\ 0 & 0 \end{array}\right)\rho_0
\,+\,{}\\
\label{eq:rho_short_time_limit}
& {}+\,\left({1\over Dt}\right)^{2\alpha}
\left(\begin{array}{c|c} 0 & -i\Delta t\tau_- \\ \hline 
-i\Delta t\tau_- & 1-2\alpha\gamma - i\epsilon t\sigma_z \end{array}\right)\rho_0\quad.
\end{align}
For the time dynamics of the Pauli matrices this gives
\begin{align}
\label{eq:sigma_x_small_times}
\langle\sigma_x\rangle(t)\,&=\,\left({1\over Dt}\right)^{2\alpha}
\left\{(1-2\alpha\gamma){\langle\sigma_x\rangle}_0 - \epsilon t{\langle\sigma_y\rangle}_0\right\}\quad,\\
\nonumber
\langle\sigma_y\rangle(t)\,&=\,\left({1\over Dt}\right)^{2\alpha}
\left\{(1-2\alpha\gamma){\langle\sigma_y\rangle}_0 + \right.\\
\label{eq:sigma_y_small_times}
&\hspace{2.5cm}
\left. +\epsilon t{\langle\sigma_x\rangle}_0
+ \Delta t{\langle\sigma_z\rangle}_0\right\}\quad,\\
\label{eq:sigma_z_small_times}
\langle\sigma_z\rangle(t)\,&=\,{\langle\sigma_z\rangle}_0 
- \left({1\over Dt}\right)^{2\alpha}\Delta t{\langle\sigma_y\rangle}_0\quad.
\end{align}
We see that this solution contains a power law arising from a resummation of all leading logarithmic terms 
$\sim(\alpha\ln(Dt))^n$, which appears also in the NIBA approximation \cite{leggett_87,weiss_12}. We note
that it is not allowed to set $t=0$ since this result is only valid for $t\gg 1/D$, i.e. terms  
$\sim\epsilon/D,\Delta/D\ll \epsilon t, \Delta t$ are neglected. 

We can study the short-time solution in two different regimes, the one for exponentially small times
$|\alpha\ln(\Omega t)|\sim 1$ where we can neglect all terms $\sim\Delta t$ and $\sim\epsilon t$, and
the one for small but not exponentially small times $|\alpha\ln(\Omega t)|\ll 1$ where 
only terms of $O(\alpha)$, $O(\alpha\ln(\Omega t))$, $O(\Delta t)$ and $O(\epsilon t)$ need to be considered.
Using (\ref{eq:tilde_delta}) we obtain for exponentially small times 
\begin{align}
\label{eq:sigma_x_exp_small_times}
\langle\sigma_x\rangle(t)\,&=\,{\tilde{\Delta}^2\over\Delta^2}\left({1\over \Omega t}\right)^{2\alpha}
(1-2\alpha\gamma){\langle\sigma_x\rangle}_0 \quad,\\
\label{eq:sigma_y_exp_small_times}
\langle\sigma_y\rangle(t)\,&=\,{\tilde{\Delta}^2\over\Delta^2}\left({1\over \Omega t}\right)^{2\alpha}
(1-2\alpha\gamma){\langle\sigma_y\rangle}_0 \quad,\\
\label{eq:sigma_z_exp_small_times}
\langle\sigma_z\rangle(t)\,&=\,{\langle\sigma_z\rangle}_0 \quad,
\end{align}
and for small but not exponentially small times
\begin{align}
\label{eq:sigma_x_not_exp_small_times}
\langle\sigma_x\rangle(t)\,&=\,{\tilde{\Delta}^2\over\Delta^2}
\left\{(1-2\alpha(\gamma+\ln(\Omega t)){\langle\sigma_x\rangle}_0 
- \epsilon t{\langle\sigma_y\rangle}_0\right\}\quad,\\
\nonumber
\langle\sigma_y\rangle(t)\,&=\,{\tilde{\Delta}^2\over\Delta^2}
\left\{(1-2\alpha(\gamma+\ln(\Omega t)){\langle\sigma_y\rangle}_0 + \right.\\
\label{eq:sigma_y_not_exp_small_times}
&\hspace{2.5cm}
\left. +\epsilon t{\langle\sigma_x\rangle}_0
+ \Delta t{\langle\sigma_z\rangle}_0\right\}\quad,\\
\label{eq:sigma_z_not_exp_small_times}
\langle\sigma_z\rangle(t)\,&=\,{\langle\sigma_z\rangle}_0 
- {\tilde{\Delta}^2\over\Delta} t{\langle\sigma_y\rangle}_0\quad.
\end{align}

For zero tunneling $\Delta=0$ the short-time solution is consistent with the exact solution
(\ref{eq:sigma_x_zero_tunneling}-\ref{eq:sigma_z_zero_tunneling}) where we set $\cos(\epsilon t)\approx 1$
and $\sin(\epsilon t)\approx \epsilon t$. In contrast, the Bloch-Redfield solution
(\ref{eq:sigma_x_small_times_Bloch_Redfield}-\ref{eq:sigma_z_small_times_Bloch_Redfield}) at small times
misses all powers of logarithmic terms $\alpha\ln(D/\Omega)$ (resummed in $\tilde{\Delta}$) and
$\alpha\ln(\Omega t)$ together with the $O(\alpha)$ corrections for 
$\langle\sigma_x\rangle(t)$ and $\langle\sigma_y\rangle(t)$. 

In the next section we will show that our analytic solution for times which are not exponentially 
small or large coincides with 
(\ref{eq:sigma_x_not_exp_small_times}-\ref{eq:sigma_z_not_exp_small_times}) in the regime of small 
but not exponentially small times. This shows that by combining the
solution (\ref{eq:sigma_x_small_times}-\ref{eq:sigma_z_small_times}) for small times with the solution
of the next section we have an analytic and systematic result one order beyond Bloch-Redfield covering the whole 
time regime from $\Omega/D\ll \Omega t\ll 1$ up to times $\Omega t\gg 1$ which are not exponentially large
(i.e. $|\alpha\ln(\Omega t)|\ll 1$).

\subsubsection{Times in the non-exponential regime}
\label{sec:non_exp_times}

We now study the regime of times which are not exponentially small or large defined by the
condition $|\alpha\ln(\Omega t)|\ll 1$. Here we can use the propagator in the form presented in 
(\ref{eq:renormalized_propagator}-\ref{eq:renormalized_sigma_a}) and apply renormalized perturbation 
theory to study the modification of the Bloch-Redfield result and to calculate the next correction
in $O(\alpha)$. We want to determine analytically the whole crossover regime from $\Omega t\ll 1$ 
up to $\Omega t\gg 1$ provided that time is not exponentially small or large such that
all logarithmic terms $\sim|\alpha\ln(\Omega t)|\ll 1$ can be treated perturbatively and 
are on the same level as terms $\sim\alpha$. In particular this
includes the long-time regime where decay sets in such that $\Gamma t\sim 1$ or
$\Omega t\sim 1/\alpha\gg 1$. In this long-time regime we have to be very careful not to expand the
resolvent $1/(E-\tilde{L}_0-\tilde{\Sigma}_a(E))$ in $\tilde{\Sigma}_a(E)\sim\alpha\Omega\sim 1/t\sim |E-z_i|$ since
$|E-z_i|$ sets the scale of the lowest order term in the denominator of the resolvent for the pole contributions.
This would be only allowed in the regime $\Omega t\lesssim 1$ but can not be used to study
the crossover to the long-time regime. Furthermore, in order to calculate systematically 
the first correction to the Bloch-Redfield result in the long-time regime $\Omega t\sim 1/\alpha$,
it is also necessary to discuss carefully terms 
$\sim\alpha^2\Omega\sim \alpha 1/t\sim \alpha|E-z_i|$ in $\tilde{\Sigma}_a(E)$. As we will see this requires
a knowledge of certain terms in $O(\alpha^2)$ of the Liouvillian but it will turn out that the contributions
of these terms to the time dynamics of $\rho(t)$ can all be related to the stationary solution up to 
$O(\alpha^2)$ which can be calculated quite efficiently in equilibrium via the partition function, 
see Ref.~\onlinecite{divincenzo_loss_05}.

To account for all these subtleties systematically we proceed as follows. Since we know in all orders
of perturbation theory that the non-analytic features of the propagator are an isolated pole at 
$E=z_{\text{st}}=0$ together with branch cuts starting at $E=z_i$, $i=0,\pm$, pointing in the direction of the 
negative imaginary axis, we can decompose the time dynamics of $\rho(t)$ in four contributions
\begin{align}
\label{eq:decomposition_time_dynamics}
\rho(t)\,=\,\rho_{\text{st}}\,+\,\sum_{i=0,\pm} \rho_i(t)\quad,
\end{align}
with
\begin{align}
\nonumber
\rho_i(t)\,&=\,{i\over 2\pi}\int_{{\cal{C}}_i} dE e^{-iEt}{1\over E-\tilde{L}_0-\tilde{\Sigma}_a(E)}\cdot\\
\label{eq:rho_i}
&\hspace{2cm}
\cdot Z'(1+\Sigma_s{1\over E})\rho_0\quad,
\end{align}
where ${\cal{C}}_i$ is a curve in the complex plane encircling clockwise the non-analytic feature 
around $E=z_i$ (i.e. an isolated pole for $i=\text{st}$ and a branch cut at $E=z_i-ix$, $x>0$, 
for $i=0,\pm$). Since the zero eigenvalue of $\tilde{L}_0+\tilde{\Sigma}_a(E)$ is projected out by the
projector $P_{\text{st}}$ given (in all orders of perturbation theory) by (\ref{eq:P_st}), we get 
for the stationary state 
\begin{align}
\nonumber
\rho_{\text{st}}\,&=\,P_{\text{st}}\rho_0 \,-\, 
{1\over \tilde{L}_0+\tilde{\Sigma}_a(0)}Z'\Sigma_s\rho_0\\
\label{eq:stationary_state}
&=\,{1\over 2}\left(\begin{array}{c} 1 \\ 1 \\ 0 \\ 0 \end{array}\right)
\,-\,{i\pi\alpha\Delta Z\over \tilde{L}_0+\tilde{\Sigma}_a(0)}{1\over 2}
\left(\begin{array}{c} 0 \\ 0 \\ 1 \\ 1 \end{array}\right)\quad,
\end{align}
where we have taken (\ref{eq:sigma_s}) and (\ref{eq:Z'_non_exp_E}) for $\Sigma_s$ and $Z'$,
respectively, and have used the normalization $\text{Tr}\rho_0=1$. For $i=0,\pm$ we obtain with
$E=-ix\pm\eta$ ($\eta=0^+$)
\begin{align}
\label{eq:rho_i_branchcut_integral}
\rho_i(t)\,=\,F_i(t) e^{-iz_it}\quad,
\end{align}
with the preexponential operator given by 
\begin{align}
\nonumber
F_i(t)\,&=\,{1\over 2\pi}\int_0^\infty \,dx \,e^{-xt}\cdot\\
\nonumber
&\hspace{-0.5cm}
\cdot\left\{{1\over E-\tilde{L}_0-\tilde{\Sigma}_a(E)}\Big|_{E=z_i-ix+\eta}\,-\,(\eta\rightarrow -\eta)\right\}\cdot\\
\label{eq:preexponential_operator}
&\hspace{1cm}
\cdot Z'(1+\Sigma_s{1\over z_i-ix})\rho_0\quad.
\end{align}
Due to the exponential part $e^{-xt}$ in (\ref{eq:preexponential_operator}), we get $x\sim 1/t$. 
The eigenvalues of $\tilde{L}_0$ are either zero or $\pm\Omega$ (see below),
and $\tilde{\Sigma}_a(E)\sim\alpha\Omega$. Thus, for times $\Omega t\lesssim 1$, $\tilde{\Sigma}_a(E)$
is a small correction in the denominator and we can expand the resolvent in $\tilde{\Sigma}_a(E)$.
However, for times $\Omega t\sim 1/\alpha$ or $|E-z_i|\sim\alpha\Omega$, 
$\tilde{\Sigma}_a(E)\sim\alpha\Omega\sim 1/t$ becomes of 
the same order as $x\sim 1/t$ and the expansion is no longer valid. To cover the crossover to this regime
as well we leave the important term $\tilde{\Sigma}_a(z_i)\sim\alpha\Omega$ in the
denominator which is essential for the correct position of the poles, and expand only in
\begin{align}
\nonumber
\tilde{\Sigma}_a(E)-\tilde{\Sigma}_a(z_i)\,&=\,\\
\nonumber & \hspace{-2cm}
=\,\alpha {\cal{F}}_i(E)M_i\,+\,\alpha\sum_{\substack{j=0,\pm \\ j\ne i}}
\left({\cal{F}}_j(E)-{\cal{F}}_j(z_i)\right)M_j\\
\nonumber & \hspace{-2cm}
\approx\,\alpha(E-z_i)\Big\{\ln{-i(E-z_i)\over\Omega}M_i + 
\sum_{\substack{j=0,\pm \\ j\ne i}}{d{\cal{F}}_j\over dE}(z_i)M_j\Big\}\\
\label{eq:delta_sigma_a} & \hspace{-2cm}
\sim\,\alpha(E-z_i)\,\sim\,{\alpha\over t}\,\ll\,{1\over t}\,\sim\,x\quad,
\end{align}
where we have used the form (\ref{eq:renormalized_sigma_a}) and $\lambda_i(z_i)=z_i$ 
(see (\ref{eq:determination_poles})), together with the fact that ${\cal{F}}_j(E)$ 
can be expanded around $E=z_i$ for $j\ne i$. 

Therefore, a systematic expansion of the resolvent up to $O(\alpha)$ valid in the whole 
non-exponential time regime is provided by 
\begin{align}
\nonumber
{1\over E-\tilde{L}_0-\tilde{\Sigma}_a(E)}&\approx {1\over E-\tilde{L}_0-\tilde{\Sigma}_a^i}\,+\\
\label{eq:resolvent_expansion}
&\hspace{-2cm}
+\,{1\over E-\tilde{L}_a^i}\delta\tilde{\Sigma}_a(E)
{1\over E-\tilde{L}_a^i}\quad,
\end{align}
where we have defined
\begin{align}
\label{eq:L_sigma_a_i}
\tilde{L}_a^i\,=\,\tilde{L}_0\,+\,\tilde{\Sigma}_a^i\quad,\quad
\tilde{\Sigma}_a^i\,=\,\tilde{\Sigma}_a(z_i)\quad,
\end{align}
and
\begin{align}
\label{eq:delta_sigma_a_i}
\delta\tilde{\Sigma}_a^i(E)\,=\,\tilde{\Sigma}_a(E)-\tilde{\Sigma}_a(z_i)\quad.
\end{align}

To complete the justification of this perturbative expansion (\ref{eq:resolvent_expansion}), we 
finally prove that the order of $E-\tilde{L}_0-\tilde{\Sigma}_a^i$ with $E=z_i-ix$ is always 
larger than $x\sim 1/t$ in the regime $\Omega t\gtrsim 1$. To show this
we denote the eigenvalues of $\tilde{L}_a^i$ by $\tilde{\gamma}_j^i$, with $i,j=\text{st},0,\pm$. 
The lowest order values are given by the eigenvalues of the real but non-hermitian Liouvillian
$\tilde{L}_0$, which can be diagonalized by the transformation 
\begin{align}
\label{eq:matrix_A}
A\,=\,A^{-1}\,=\,
{1\over \Omega}
\left(\begin{array}{cc} \Omega\tau_++\epsilon\tau_- & -\Delta\sigma_z\tau_+ \\ 
-\Delta Z\sigma_z\tau_- & -\epsilon\tau_+-\Omega\tau_- \end{array}\right)\quad,
\end{align}
which is the analog of (\ref{eq:matrix_A_0}) but with $\Omega_0\rightarrow\Omega$ and the $Z$-factor 
in the lower non-diagonal resulting in a non-unitary matrix. In this basis $\tilde{L}_0$ is given by
\begin{align}
\label{eq:tilde_L_0_transformed}
A \tilde{L}_0 A \,=\, 
\left(\begin{array}{cc} 0 & 0 \\ 0 & \Omega\sigma_z \end{array}\right)\quad,
\end{align}
i.e. two eigenvalues are zero and two are identical to $\pm\Omega$ in lowest order in $\alpha$. 
$\tilde{\Sigma}_a^i$ will shift these eigenvalues by $O(\alpha\Omega)$ such that, together with 
the symmetry relations (\ref{eq:eigenvalue_st}-\ref{eq:eigenvalue_sigma}), we get 
\begin{align}
\label{eq:gamma_poles}
\tilde{\gamma}_{\text{st}}^i &= 0\quad,\quad
\tilde{\gamma}_0^0 = z_0 \quad,\quad
\tilde{\gamma}_\sigma^\sigma = z_\sigma \quad,\\
\label{eq:gamma_0_sigma}
\tilde{\gamma}_0^\sigma &= -\left(\tilde{\gamma}_0^{-\sigma}\right)^* = O(\alpha\Omega)\quad,\\
\label{eq:gamma_sigma_-sigma}
\tilde{\gamma}_{-\sigma}^\sigma &= -\left(\tilde{\gamma}_\sigma^{-\sigma}\right)^* 
= -\sigma\Omega + O(\alpha\Omega)\quad,\\
\label{eq:gamma_sigma_0}
\tilde{\gamma}_{\sigma}^0 &= -\left(\tilde{\gamma}_{-\sigma}^0\right)^*
= \sigma\Omega + O(\alpha\Omega)\quad,
\end{align}
We note that (\ref{eq:gamma_poles}) holds exactly in all orders of perturbation theory since
$1/(E-L_a(E))=(1/(E-\tilde{L}_a(E)))Z'$ with $Z'$ given by (\ref{eq:Z'_non_exp_E}) can be viewed as 
the definition of $\tilde{L}_a(E)$ and, therefore, the pole positions of the two resolvents 
$1/(E-L_a(E))$ and $1/(E-\tilde{L}_a(E))$ must be exactly the same. 

For $E=z_i-ix$, (\ref{eq:gamma_poles}-\ref{eq:gamma_sigma_0}) leads to 
\begin{align}
\label{eq:eigenvalues_tilde_L_a}
|E-\tilde{\gamma}_j^i|\,=\,
\begin{cases}
x
& \text{for $j=i$} \\
|-ix+z_i|
& \text{for $j=\text{st}$} \\
|-ix\pm\Omega + O(\alpha\Omega)|
& \text{for $j\ne i,\text{st}$}
\end{cases}
\end{align}
i.e. for $x\sim 1/t\lesssim\Omega$ to the desired result 
\begin{align}
\label{eq:perturbation_expansion_condition}
|E-\tilde{\gamma}_j^i|\,\gtrsim\,x\quad.
\end{align}

Using the expansion (\ref{eq:resolvent_expansion}) it is now straightforward to write down
the various terms for the time dynamics of $\rho_i(t)$. Denoting the projectors on the eigenstates
of $\tilde{L}_0+\tilde{\Sigma}_a^i$ by $\tilde{P}_j^i$, with
$i,j=\text{st},0,\pm$, we get
for $i=0,\pm$
\begin{align}
\nonumber
\rho_i(t)\,&=\,{i\over 2\pi}\int_{{\cal{C}}_i} dE e^{-iEt}
\Big\{{1\over E-\tilde{\gamma}_i^i}\tilde{P}_i^i \,+\,\\
\nonumber & 
+\,\sum_{j,j'=0,\pm}{1\over E-\tilde{\gamma}_j^i}\,{1\over E-\tilde{\gamma}_{j'}^i}
\tilde{P}_j^i\delta\tilde{\Sigma}_a^i(E)\tilde{P}_{j'}^i\Big\}\cdot\\
\label{eq:rho_i_perturbative}
&\hspace{2cm}
\cdot Z'(1+\Sigma_s{1\over E})\rho_0\quad.
\end{align}
Here, we have used for the first term (in the first bracket) on the r.h.s. 
that the other projectors $\tilde{P}_j^i$ with $j\ne i$ lead to an
analytic function on the curve ${\cal{C}}_i$ with zero integral. Furthermore, due to the matrix 
structure (\ref{eq:delta_sigma_a}) of $\delta\tilde{\Sigma}_a^i(E)$ and the form (\ref{eq:P_st}) of 
$\tilde{P}_{\text{st}}^i$, we get 
$\tilde{P}_{\text{st}}^i\delta\tilde{\Sigma}_a^i(E)=\delta\tilde{\Sigma}_a^i(E)\tilde{P}_{\text{st}}^i=0$ 
and only the terms with $j,j'\ne\text{st}$ contribute to the second term (in the first bracket) on the
r.h.s. Furthermore, we note that we can omit all analytic terms
$\sim (E-z_i)^2$ for $\delta\tilde{\Sigma}_a^i(E)$ since they lead to analytic contributions on the 
curve ${\cal{C}}_i$ in (\ref{eq:rho_i_perturbative}) with zero integral. Thus, we  
can use the form (\ref{eq:delta_sigma_a}) for $\delta\tilde{\Sigma}_a^i(E)$. 
Inserting this form and leaving out all analytic
functions on ${\cal{C}}_i$, we can split $\rho_i(t)$ obviously in pole and pure branch cut contributions
\begin{align}
\label{eq:rho_i_pole_bc}
\rho_i(t)\,=\,\rho_i^p(t)\,+\,\rho_i^{\text{bc}}(t)\quad,
\end{align}
with 
\begin{align}
\label{eq:rho_i_pole_decomposition}
\rho_i^p(t)\,&=\,\rho_i^{p1}(t)\,+\,\rho_i^{p2}(t)\,+\,\rho_i^{p3}(t)\quad,\\
\label{eq:rho_i_pole_1}
\rho_i^{p1}(t)\,&=\,\tilde{P}_i^i Z'(1+\Sigma_s{1\over z_i})\rho_0 \,e^{-iz_it}\quad,\\
\nonumber
\rho_i^{p2}(t)\,&=\,\alpha \sum_{\substack{j=0,\pm\\j\ne i}}{d{\cal{F}}_j\over dE}(z_i)
\tilde{P}_i^i M_j \tilde{P}_i^i\cdot\\
\label{eq:rho_i_pole_2}
&\hspace{2cm}
\cdot Z'(1+\Sigma_s{1\over z_i})\rho_0\,e^{-iz_it}\quad,\\
\nonumber
\rho_i^{p3}(t)\,&=\,\alpha
{i\over 2\pi}\int_{{\cal{C}}_i} dE e^{-iEt}{1\over E-z_i}\ln{-i(E-z_i)\over\Omega}\cdot\\
\label{eq:rho_i_pole_3}
&\hspace{2cm}
\cdot \tilde{P}_i^i M_i \tilde{P}_i^i Z'(1+\Sigma_s{1\over E})\rho_0
\end{align}
for the pole contributions and
\begin{align}
\nonumber
\rho_i^{bc}(t)\,&=\,\alpha {i\over 2\pi}\int_{{\cal{C}}_i} dE e^{-iEt}
(E-z_i)\ln{-i(E-z_i)\over\Omega}\cdot\\
\label{eq:rho_i_bc} 
&\hspace{-1cm}
\cdot \sum_{\substack{j,j'=0,\pm\\(j,j')\ne(i,i)}}
{1\over (E-\tilde{\gamma}_j^i)(E-\tilde{\gamma}_{j'}^i)}
\tilde{P}_j^i M_i \tilde{P}_{j'}^i Z'(1+\Sigma_s{1\over E})\rho_0
\end{align}
for the pure branch cut contributions. We note that the terms involving $\Sigma_s/z_i$ are very
important for (\ref{eq:rho_i_pole_1}) to calculate the terms in $O(1)$ and $O(\alpha)$ 
consistently since
\begin{align}
\label{eq:sigma_s_z_0}
\Sigma_s{1\over z_0}\,&=\,-{\Delta\Omega\over\tilde{\Delta}^2}\Big(1-{\Gamma^{(2)}\over\Gamma^{(1)}}\Big)
\left(\begin{array}{cc} 0 & 0 \\ \tau_+ & 0  \end{array}\right)\,+\,O(\alpha^2)\quad,\\
\label{eq:sigma_s_z_sigma}
\Sigma_s{1\over z_\sigma}\,&=\, i\sigma\pi\alpha{\Delta\over\Omega}
\left(\begin{array}{cc} 0 & 0 \\ \tau_+ & 0  \end{array}\right)\,+\,O(\alpha^2)\quad,
\end{align}
where we have used (\ref{eq:sigma_s}) for $\Sigma_s$ and expanded the pole position $z_0=-i\Gamma$
in $\alpha$ by using
\begin{align}
\label{eq:expansion_gamma}
\Gamma\,=\,\Gamma^{(1)}+\Gamma^{(2)}+O(\alpha^3)\quad,\quad 
\Gamma^{(1)}\,=\,\pi\alpha{\tilde{\Delta}^2\over\Omega}\quad,
\end{align}
where we have taken (\ref{eq:gamma_RG}) for $\Gamma^{(1)}$. This shows that also second order terms
$\sim\alpha^2$ are needed for the Liouvillian to calculate the pole position $z_0$ up to second order
needed to get all terms in $O(\alpha)$ for the purely decaying mode of the time evolution. A similiar 
term will also occur for the stationary state (see below). 

In contrast, for the two other pole contributions (\ref{eq:rho_i_pole_2}) and (\ref{eq:rho_i_pole_3})
the terms involving $\Sigma_s$ are only needed for the purely decaying mode $i=0$ and it is sufficient to
take $z_0$ up to $O(\alpha)$. For the branch cut contribution (\ref{eq:rho_i_bc}) the term with $\Sigma_s$ 
can be left out since it leads to a contribution in $O(\alpha^2)$. Furthermore, in 
(\ref{eq:rho_i_pole_2}-\ref{eq:rho_i_bc}) the projectors $\tilde{P}_j^i$ and
the eigenvalues $\tilde{\gamma}_j^i$ can be
replaced by their values in lowest order, all other terms contribute in $O(\alpha^2)$. Only for 
the first pole contribution (\ref{eq:rho_i_pole_1}) the projector $\tilde{P}_i^i$ is needed up to $O(\alpha)$.
Denoting the projectors in lowest and in first order in $\alpha$ by
$\tilde{P}^{(0)i}_j$ and $\tilde{P}^{(1)i}_j$, we show in Appendix~\ref{sec:appendix_projectors_tilde_L_a}
by a straightforward calculation that the projectors transformed with the matrix $A$ 
(see (\ref{eq:matrix_A})) are given by
\begin{align}
\label{eq:P_0_i_0}
A \tilde{P}^{(0)i}_0 A\,&=\,
\left(\begin{array}{cc} \tau_- & 0 \\ 0 & 0  \end{array}\right)\quad,\\
\label{eq:P_0_i_sigma}
A \tilde{P}^{(0)i}_\sigma A\,&=\,{1\over 2}
\left(\begin{array}{cc} 0 & 0 \\ 0 & 1+\sigma\sigma_z \end{array}\right)\quad,\\
\label{eq:P_1_0_0}
A \tilde{P}^{(1)0}_0 A\,&=\,i\pi\alpha{\tilde{\Delta}^2\epsilon\over\Delta\Omega^2}
\left(\begin{array}{cc} 0 & {1\over Z}\tau_- \\ \tau_- & 0  \end{array}\right)\quad,\\
\label{eq:P_1_sigma_sigma}
A \tilde{P}^{(1)\sigma}_\sigma A\,&=\,{1\over 4}i\pi\sigma\alpha{\tilde{\Delta}^2\over\Omega^2}
\left(\begin{array}{cc} 0 & 0 \\ 0 & \tau_+-\tau_- \end{array}\right)\quad,
\end{align}
Taking the projectors in lowest order, the number of terms contributing to 
(\ref{eq:rho_i_pole_2}-\ref{eq:rho_i_bc}) is considerably reduced due to 
\begin{align}
\label{eq:PMP_0_combinations}
\tilde{P}^{(0)0}_j M_0 \tilde{P}^{(0)0}_{j'} &\ne 0 \Leftrightarrow j,j'\ne 0 \quad, \\
\label{eq:PMP_sigma_combinations}
\tilde{P}^{(0)\sigma}_j M_\sigma \tilde{P}^{(0)\sigma}_{j'} &\ne 0 \Leftrightarrow j,j'\ne -\sigma \quad,\\
\label{eq:PMP_special_combination}
\tilde{P}^{(0)\sigma}_\sigma M_{-\sigma} \tilde{P}^{(0)\sigma}_\sigma &= 0 \quad.\\
\end{align}
As a consequence we get
\begin{align}
\nonumber
\rho_0^{p1}(t)\,&=\,\Big\{\tilde{P}^{(0)0}_0 Z'\Big(1+i\Sigma_s{1\over\Gamma^{(1)}}
(1-{\Gamma^{(2)}\over\Gamma^{(1)}})\Big)\\
\label{eq:rho_0_pole_1}
&\hspace{1cm}
+ \tilde{P}^{(1)0}_0 Z'\Big\}\rho_0 \,e^{-iz_0t}\quad,\\
\nonumber
\rho_0^{p2}(t)\,&=\,\alpha \sum_{\sigma=\pm}{d{\cal{F}}_\sigma\over dE}(z_0)
\tilde{P}^{(0)0}_0 M_\sigma \tilde{P}^{(0)0}_0\cdot\\
\label{eq:rho_0_pole_2}
&\hspace{2cm}
\cdot Z'(1+i\Sigma_s{1\over \Gamma^{(1)}})\rho_0\,e^{-iz_0t}\quad,\\
\label{eq:rho_0_pole_3}
\rho_0^{p3}(t)\,&=\,0\quad,\\
\nonumber
\rho_0^{bc}(t)\,&=\,\alpha\sum_{\sigma,\sigma'=\pm}
{i\over 2\pi}\int_{{\cal{C}}_0} dE e^{-iEt}\ln{-i(E-z_0)\over\Omega}\cdot\\
\label{eq:rho_0_bc} 
&\hspace{0cm}
\cdot {(E-z_0)\over (E-\sigma\Omega)(E-\sigma'\Omega)}
\tilde{P}^{(0)0}_\sigma M_0 \tilde{P}^{(0)0}_{\sigma'} Z'\rho_0\,\,,
\end{align}
and
\begin{align}
\nonumber
\rho_\sigma^{p1}(t)\,&=\,\Big\{\tilde{P}^{(0)\sigma}_\sigma
Z'(1+{\sigma\over\Omega}\Sigma_s) + \\
\label{eq:rho_sigma_pole_1}
&\hspace{1cm}
+ \tilde{P}^{(1)\sigma}_\sigma Z'\Big\}\rho_0 \,e^{-iz_\sigma t}\quad,\\
\nonumber
\rho_\sigma^{p2}(t)\,&=\,\alpha {d{\cal{F}}_0 \over dE}(z_\sigma)
\tilde{P}^{(0)\sigma}_\sigma M_0 \tilde{P}^{(0)\sigma}_\sigma\cdot\\
\label{eq:rho_sigma_pole_2}
&\hspace{2cm}
\cdot Z'(1+i\Sigma_s{1\over \Gamma^{(1)}})\rho_0\,e^{-iz_\sigma t}\quad,\\
\nonumber
\rho_\sigma^{p3}(t)\,&=\,\alpha
{i\over 2\pi}\int_{{\cal{C}}_\sigma} dE e^{-iEt}{1\over E-z_\sigma}\ln{-i(E-z_\sigma)\over\Omega}\cdot\\
\label{eq:rho_sigma_pole_3}
&\hspace{2cm}
\cdot \tilde{P}^{(0)\sigma}_\sigma M_\sigma \tilde{P}^{(0)\sigma}_\sigma Z'\rho_0\quad,\\
\nonumber
\rho_\sigma^{bc}(t)\,&=\,\alpha {i\over 2\pi}\int_{{\cal{C}}_\sigma} dE e^{-iEt}\ln{-i(E-z_\sigma)\over\Omega}\cdot\\
\nonumber
&\hspace{-1cm}
\cdot \Big\{{E-z_\sigma\over E^2}\tilde{P}^{(0)\sigma}_0 M_\sigma \tilde{P}^{(0)\sigma}_0 + \\
\label{eq:rho_sigma_bc} 
&\hspace{-0.5cm}
+{1\over E}(\tilde{P}^{(0)\sigma}_0 M_\sigma \tilde{P}^{(0)\sigma}_\sigma + 
\tilde{P}^{(0)\sigma}_\sigma M_\sigma \tilde{P}^{(0)\sigma}_0) \Big\}
Z'\rho_0\,\,.
\end{align}

The first term on the r.h.s. of (\ref{eq:rho_0_pole_1}) and (\ref{eq:rho_sigma_pole_1}) leads 
to the Bloch-Redfield result modified by the $Z$-factor. All other contributions to the time
evolution are corrections in $O(\alpha)$. All energy integrals can be calculated from
\begin{align}
\nonumber
&{i\over 2\pi}\int dE e^{-iEt}{1\over E-z_i}\ln{-i(E-z_i)\over\Omega}\,=\\
\label{eq:integral_ln}
&\hspace{1cm}
=\,-(\gamma + \ln(\Omega t))e^{-iz_i t}\\
\nonumber
&{i\over 2\pi}\int_{{\cal{C}}_i} dE e^{-iEt}{1\over E-a}\ln{-i(E-z_i)\over\Omega}\,=\\
\nonumber
&\hspace{1cm}
=\,e^{-iz_i t}\int_0^\infty dy e^{-y}{1\over y-i(a-z_i)t}\,=\\
\label{eq:integral_H}
&\hspace{1cm}
=\,e^{-iz_i t}H((a-z_i)t)\quad,\\
\nonumber
&{i\over 2\pi}\int_{{\cal{C}}_i}dE e^{-iEt}{E-z_i\over (E-a)^2}\ln{-i(E-z_i)\over\Omega}\,=\\
\nonumber
&\hspace{1cm}
=\,e^{-iz_i t}\int_0^\infty dy e^{-y}{y\over (y-i(a-z_i)t)^2}\,=\\
\label{eq:integral_tilde_H}
&\hspace{1cm}
=\,e^{-iz_i t}\tilde{H}((a-z_i)t)\quad,
\end{align}
where $\gamma$ is Euler's constant, $a\ne z_i$, and $H(x)$ and $\tilde{H}(x)$ 
can be expressed via the exponential integral $E_1(-ix)$ 
\begin{align}
\label{eq:H_function}
H(x)\,&=\,e^{-ix}E_1(-ix)\quad,\\
\label{eq:tilde_H_function}
\tilde{H}(x)\,&=\,(1-ix)H(x)-1\quad.
\end{align}
It is important to note that, for the energy integrals occurring in (\ref{eq:rho_sigma_bc}) and 
(\ref{eq:rho_0_bc}), the imaginary part of $(a-z_i)t$ is $\sim -i\Gamma t$ and can be neglected
in $H((a-z_i)t)$ and $\tilde{H}((a-z_i)t)$ (i.e. leading to higher orders in $\alpha$) compared to 
the real part of $(a-z_i)t$ which is given by $\pm \Omega t$. This
holds even in the case $\Gamma t\sim 1$, as can be seen from the integrals (\ref{eq:integral_H}) and
(\ref{eq:integral_tilde_H}). In contrast, for the exponential function $e^{-iz_i t}$ it is not possible
to expand in the imaginary part of $z_i$ for $\Gamma t\sim 1$. As a consequence, only the crossover
functions $H(\pm\Omega t)$ and $\tilde{H}(\pm\Omega t)$ will appear for the branch cut integrals.

Finally, the derivatives of ${\cal{F}}_i(E)$ can be obtained from (\ref{eq:F})
\begin{align}
\label{eq:derivative_F}
{d {\cal{F}}_i\over dE}(E)\,=\,1+\ln{-i(E-\lambda_i(E))\over\Omega}+O(\alpha)\quad,  
\end{align}
which gives
\begin{align}
\label{eq:derivative_F_0}
{d {\cal{F}}_0\over dE}(z_\sigma)\,&=\,1-i\sigma{\pi\over 2} + O(\alpha)\quad,\\
\label{eq:derivative_F_sigma}
{d {\cal{F}}_\sigma\over dE}(z_0)\,&=\,1+i\sigma{\pi\over 2} + O(\alpha)\quad.
\end{align}

Using all these relationships together with the form of the various matrices, one can
straightforwardly evaluate (\ref{eq:rho_0_pole_1}-\ref{eq:rho_sigma_bc}) and calculate the expectation
values of the Pauli matrices. Decomposing the time dynamics according to 
(\ref{eq:time_dynamics_pauli_general_form}) in the various modes, we get for the preexponential functions the
following final result for the time dynamics in the non-exponential time regime 
\begin{widetext}
\begin{align}
\label{eq:F_x_0_full}
F_x^0(t) \,&=\,
-{\langle\sigma_x\rangle}_{\text{st}} 
\,-\, \Big(1+2\alpha{\tilde{\Delta}^2\over\Omega^2}\Big){\tilde{\Delta}^2\over\Delta\Omega}
{\langle\tilde{\sigma}_z\rangle}_0
\,+\, \alpha{\tilde{\Delta}^4\epsilon\over\Delta^2\Omega^3}
\Big\{{(H_t^\prime - \tilde{H}_t^\prime)\langle\tilde{\sigma}_x\rangle}_0 + 
(\pi + \tilde{H}_t^{\prime\prime}){\langle\sigma_y\rangle}_0\Big\}
\quad,\\
\label{eq:F_y_0_full}
F_y^0(t) \,&=\,
\pi\alpha{\tilde{\Delta}^2\epsilon\over\Delta\Omega^2}(1+{\langle\tilde{\sigma}_z\rangle}_0)
\,+\, \alpha{\tilde{\Delta}^4\over\Delta^2\Omega^2}
\Big\{\tilde{H}_t^{\prime\prime}{\langle\tilde{\sigma}_x\rangle}_0+
(H_t^\prime + \tilde{H}_t^\prime){\langle\sigma_y\rangle}_0\Big\}
\quad,\\
\label{eq:F_z_0_full}
F_z^0(t) \,&=\,
-{\langle\sigma_z\rangle}_{\text{st}}
\,+\, \Big(1+2\alpha{\tilde{\Delta}^2\over\Omega^2}\Big){\epsilon\over\Omega}{\langle\tilde{\sigma}_z\rangle}_0
\,-\, \pi\alpha{\tilde{\Delta}^2\epsilon^2\over\Delta\Omega^3}{\langle\sigma_y\rangle}_0
\,+\, \alpha{\tilde{\Delta}^4\over\Delta\Omega^3}
\Big\{(H_t^\prime - \tilde{H}_t^\prime){\langle\tilde{\sigma}_x\rangle}_0 + 
\tilde{H}_t^{\prime\prime}{\langle\sigma_y\rangle}_0\Big\}
\quad,\\
\label{eq:F_x_c_full}
F_x^c(t) \,&=\,
-f_t{\tilde{\Delta}^2\epsilon\over\Delta^2\Omega}{\langle\tilde{\sigma}_x\rangle}_0
\,-\, \alpha{\tilde{\Delta}^4\epsilon\over\Delta^2\Omega^3}
\Big\{2 H_t^\prime{\langle\tilde{\sigma}_x\rangle}_0 - 
({\pi\over 2} + 2 H_t^{\prime\prime}){\langle\sigma_y\rangle}_0\Big\}
\,-\, 2\alpha{\tilde{\Delta}^2\over\Delta\Omega^3}
(\tilde{\Delta}^2 \tilde{H}_t^\prime + \epsilon^2 H_t^\prime){\langle\tilde{\sigma}_z\rangle}_0 
\quad,\\
\label{eq:F_y_c_full}
F_y^c(t) \,&=\,
f_t{\tilde{\Delta}^2\over\Delta^2}{\langle\sigma_y\rangle}_0
\,-\, \alpha{\tilde{\Delta}^2\epsilon\over\Delta\Omega^2}(\pi + 2 H_t^\prime{\langle\tilde{\sigma}_z\rangle}_0)
\,-\, {\pi\over 2}\alpha{\tilde{\Delta}^4\over\Delta^2\Omega^2}{\langle\tilde{\sigma}_x\rangle}_0
\quad,\\
\label{eq:F_z_c_full}
F_z^c(t) \,&=\,
-f_t{\tilde{\Delta}^2\over\Delta\Omega}{\langle\tilde{\sigma}_x\rangle}_0
\,-\,{\pi\over 2}\alpha{\tilde{\Delta}^4\over\Delta\Omega^3}{\langle\sigma_y\rangle}_0
\,+\, 2\alpha{\tilde{\Delta}^2\epsilon^2\over\Delta\Omega^3}
( H_t^\prime{\langle\tilde{\sigma}_x\rangle}_0 + H_t^{\prime\prime}{\langle\sigma_y\rangle}_0)
\,-\, 2\alpha{\tilde{\Delta}^2\epsilon\over\Omega^3}
(H_t^\prime-\tilde{H}_t^\prime){\langle\tilde{\sigma}_z\rangle}_0
\quad,\\
\label{eq:F_x_s_full}
F_x^s(t) \,&=\,
-f_t{\tilde{\Delta}^2\epsilon\over\Delta^2\Omega}{\langle\sigma_y\rangle}_0
\,+\, \pi\alpha{\tilde{\Delta}^2\epsilon^2\over\Delta\Omega^3}
\,+\, 2\alpha{\tilde{\Delta}^4\epsilon\over\Delta^2\Omega^3}
( H_t^{\prime\prime}{\langle\tilde{\sigma}_x\rangle}_0- H_t^\prime{\langle\sigma_y\rangle}_0)
\,+\, 2\alpha{\tilde{\Delta}^2\over\Delta\Omega^3}{
(\epsilon^2 H_t^{\prime\prime}+\tilde{\Delta}^2\tilde{H}_t^{\prime\prime})\langle\tilde{\sigma}_z\rangle}_0
\quad,\\
\label{eq:F_y_s_full}
F_y^s(t) \,&=\,
-f_t{\tilde{\Delta}^2\over\Delta^2}{\langle\tilde{\sigma}_x\rangle}_0
\,-\, \pi\alpha{\tilde{\Delta}^4\over\Delta^2\Omega^2}{\langle\sigma_y\rangle}_0
\,-\, 2\alpha{\tilde{\Delta}^2\epsilon\over\Delta\Omega^2} H_t^\prime{\langle\tilde{\sigma}_z\rangle}_0
\quad,\\
\label{eq:F_z_s_full}
F_z^s(t) \,&=\,
-f_t{\tilde{\Delta}^2\over\Delta\Omega}{\langle\sigma_y\rangle}_0
\,+\, \alpha{\tilde{\Delta}^2\epsilon\over\Omega^3}
\Big\{\pi + 2(H_t^{\prime\prime}-\tilde{H}_t^{\prime\prime}){\langle\tilde{\sigma}_z\rangle}_0\Big\}
\,-\, 2\alpha{\tilde{\Delta}^2\epsilon^2\over\Delta\Omega^3}
(H_t^{\prime\prime}{\langle\tilde{\sigma}_x\rangle}_0 - H_t^{\prime}{\langle\sigma_y\rangle}_0)
\quad,
\end{align}
\end{widetext}
where we have defined the quantities
\begin{align}
\label{eq:H_prime_t}
H_t^\prime\,&=\,\text{Re}H(\Omega t)\,=\,{1\over 2}\sum_{\sigma=\pm}H(\sigma\Omega t)\quad,\\
\label{eq:H_prime_prime_t}
H_t^{\prime\prime}\,&=\,\text{Im}H(\Omega t)\,=\,-{i\over 2}\sum_{\sigma=\pm}\sigma H(\sigma\Omega t)\quad,\\
\label{eq:tilde_H_prime_t}
\tilde{H}_t^\prime\,&=\,\text{Re}\tilde{H}(\Omega t)\,=\,{1\over 2}\sum_{\sigma=\pm}\tilde{H}(\sigma\Omega t)\quad,\\
\label{eq:tilde_H_prime_prime_t}
\tilde{H}_t^{\prime\prime}\,&=\,\text{Im}\tilde{H}(\Omega t)
\,=\,-{i\over 2}\sum_{\sigma=\pm}\sigma \tilde{H}(\sigma\Omega t)\quad,\\
\label{eq:tilde_sigma_x_renormalized}
\tilde{\sigma}_x\,&=\,
-{1\over\Omega}\Big(\epsilon\sigma_x + \Delta\sigma_z\Big)\quad,\\
\label{eq:tilde_sigma_z_renormalized}
\tilde{\sigma}_z\,&=\,{1\over\Omega}\Big(\epsilon\sigma_z - 
{\tilde{\Delta}^2\over\Delta}\sigma_x\Big)\quad,\\
\label{eq:f_t}
f_t\,&=\,1 + \alpha{\tilde{\Delta}^2\over\Omega^2} 
- 2\alpha\Big(\gamma + \ln(\Omega t)\Big){\epsilon^2\over\Omega^2}\quad.
\end{align}
We note that the operators $\tilde{\sigma}_x$ and $\tilde{\sigma}_z$ can not be interpreted as
the Pauli spin operators in the basis where the local Hamiltonian with
$\Delta\rightarrow\tilde{\Delta}$ is diagonal since {\it both} $\Delta$ and $\tilde{\Delta}$ appear
in the definition in a subtle way. Only if the renormalization of the tunneling is neglected, 
these operators are identical to the Pauli spin operators defined in 
(\ref{eq:sigma_x_prime}-\ref{eq:sigma_z_prime}).

The stationary values ${\langle\sigma_\alpha\rangle}_{\text{st}}$ of the Pauli matrices follow from
\begin{align}
\label{eq:sigma_x_st_full}
{\langle\sigma_x\rangle}_{\text{st}}\,&=\,{\tilde{\Delta}^2\over\Delta\Omega}
\Big(1-{\Gamma^{(2)}\over\Gamma^{(1)}}\Big) + 2\alpha{\tilde{\Delta}^4\over\Delta\Omega^3}\quad,\\
\label{eq:sigma_y_st_full}
{\langle\sigma_y\rangle}_{\text{st}}\,&=\,0\quad,\\
\label{eq:sigma_z_st_full}
{\langle\sigma_z\rangle}_{\text{st}}\,&=\,-{\epsilon\over\Omega}
\Big(1-{\Gamma^{(2)}\over\Gamma^{(1)}}\Big) - 2\alpha{\tilde{\Delta}^2\epsilon\over\Omega^3}\quad.
\end{align}
This can be obtained from (\ref{eq:stationary_state}) via the spectral decomposition of 
$\tilde{L}_a^{\text{st}}=\tilde{L}_0+\tilde{\Sigma}_a(0)$. Denoting the eigenvalues and projectors of this
Liouvillian by $\tilde{\gamma}_j^{\text{st}}$ and $\tilde{P}_j^{\text{st}}$, with $j=\text{st},0,\pm$, we show in 
Appendix~\ref{sec:appendix_projectors_tilde_L_a} that we get in analogy to 
(\ref{eq:gamma_poles}-\ref{eq:gamma_sigma_0}) and (\ref{eq:P_0_i_0}-\ref{eq:P_1_sigma_sigma})
\begin{align}
\label{eq:gamma_st_st}
\tilde{\gamma}_{\text{st}}^{\text{st}} &= 0\quad,\\
\label{eq:gamma_0_st}
\tilde{\gamma}_0^{\text{st}} &= 
-i\Big(\Gamma^{(1)}(1-2\alpha{\tilde{\Delta}^2\over\Omega^2})+\Gamma^{(2)}\Big)+O(\alpha^3)\quad,\\
\label{eq:gamma_sigma_st}
\tilde{\gamma}_{\sigma}^{\text{st}} &= \sigma\Omega + O(\alpha\Omega)\quad,
\end{align}
and 
\begin{align}
\label{eq:P_st_st}
A \tilde{P}^{\text{st}}_{\text{st}} A\,&=\,
\left(\begin{array}{cc} \tau_+ & 0 \\ 0 & 0  \end{array}\right)\quad,\\
\label{eq:P_0_st_0}
A \tilde{P}^{(0)\text{st}}_0 A\,&=\,
\left(\begin{array}{cc} \tau_- & 0 \\ 0 & 0  \end{array}\right)\quad,\\
\label{eq:P_0_st_sigma}
A \tilde{P}^{(0)\text{st}}_\sigma A\,&=\,{1\over 2}
\left(\begin{array}{cc} 0 & 0 \\ 0 & 1+\sigma\sigma_z \end{array}\right)\quad,\\
\label{eq:P_1_st_0}
A \tilde{P}^{(1)\text{st}}_0 A\,&=\,i\pi\alpha{\tilde{\Delta}^2\epsilon\over\Delta\Omega^2}
\left(\begin{array}{cc} 0 & {1\over Z}\tau_- \\ \tau_- & 0  \end{array}\right)\quad.
\end{align}
Inserting the spectral decomposition in (\ref{eq:stationary_state}) we get up to $O(\alpha)$
\begin{align}
\nonumber
\rho_{\text{st}}\,&=\,
{1\over 2}\left(\begin{array}{c} 1 \\ 1 \\ 0 \\ 0 \end{array}\right)
-{i\Gamma^{(1)}\over 2\Delta}
\Big\{{\Omega\over\tilde{\gamma}_0^{\text{st}}}\tilde{P}_0^{(0)\text{st}}+\\
&\hspace{0cm}
+i{\Omega\over\Gamma^{(1)}}\tilde{P}_0^{(1)\text{st}}
+\sum_{\sigma=\pm}\sigma \tilde{P}_\sigma^{(0)\text{st}}\Big\}
\left(\begin{array}{c} 0 \\ 0 \\ 1 \\ 1 \end{array}\right)\quad.
\end{align}
Inserting (\ref{eq:gamma_0_st}) and (\ref{eq:P_0_st_0}-\ref{eq:P_1_st_0}) we find that the sum of the
last two terms on the r.h.s. is zero and we get for the stationary density matrix up to $O(\alpha)$
the final result
\begin{align}
\label{eq:rho_stationary_full}
\rho_{\text{st}}\,&=\,
{1\over 2}\left(\begin{array}{c} 1 \\ 1 \\ 0 \\ 0 \end{array}\right)-
\Big(1+2\alpha{\tilde{\Delta}^2\over\Omega^2}-{\Gamma^{(2)}\over\Gamma^{(1)}}\Big){1\over 2\Omega}
\left(\begin{array}{c} \epsilon \\ -\epsilon \\ 
-\tilde{\Delta}^2/\Delta \\ -\tilde{\Delta}^2/\Delta \end{array}\right)\quad,
\end{align}
which leads to the result (\ref{eq:sigma_x_st_full}-\ref{eq:sigma_z_st_full}) for the stationary values of
the Pauli matrices.

To calculate the ratio $\Gamma^{(2)}/\Gamma^{(1)}$ we need an analysis of all second order terms 
$\sim\alpha^2$ of the Liouvillian $L(0^+)$ to get the stationary state up to $O(\alpha)$. This
goes beyond the scope of this paper. However, in Ref.~\onlinecite{divincenzo_loss_05}, such an
analysis has been performed in bare perturbation theory (i.e. using the unrenormalized tunneling) 
with the result (note that we slightly changed the result such that it is valid for a Lorentzian
cutoff function in the bath)
\begin{align}
\label{eq:sigma_x_unrenormalized}
{\langle\sigma_x\rangle}_{\text{st}}\,&=\,{\Delta\over\Omega_0}+\alpha{\Delta^3\over\Omega^3}
+\alpha{\Delta\over\Omega_0^3}(\Delta^2+2\epsilon^2)\ln{\Omega_0\over D}\quad.
\end{align}
and it was shown that this agrees with the result from the partition function proving the Ergoden
hypothesis up to $O(\alpha)$. This result is consistent with (\ref{eq:sigma_x_st_full}) if we take
\begin{align}
\label{eq:ratio_Gamma2_Gamma1}
{\Gamma^{(2)}\over\Gamma^{(1)}}\,=\,\alpha{\tilde{\Delta}^2\over\Omega^2}\quad,
\end{align}
such that our final result for the stationary values reads
\begin{align}
\label{eq:sigma_x_st_full_final}
{\langle\sigma_x\rangle}_{\text{st}}\,&=\,{\tilde{\Delta}^2\over\Delta\Omega}
+\alpha{\tilde{\Delta}^4\over\Delta\Omega^3}\quad,\\
\label{eq:sigma_y_st_full_final}
{\langle\sigma_y\rangle}_{\text{st}}\,&=\,0\quad,\\
\label{eq:sigma_z_st_full_final}
{\langle\sigma_z\rangle}_{\text{st}}\,&=\,-{\epsilon\over\Omega}
-\alpha{\tilde{\Delta}^2\epsilon\over\Omega^3}\quad.
\end{align}

We note that the terms involving $\Gamma^{(2)}$ cancel out for the full time dynamics of $\rho(t)$
in the limit $\Gamma t\ll 1$, where the exponential $e^{-\Gamma t}\approx 1$. This is a generic feature
since, in this time regime, $|E-L_0|\gg\Gamma$, and bare perturbation theory can be used to expand
the resolvent $1/(E-L_0-\Sigma(E))$ in $\Sigma(E)$, without any need of the Liouvillian up to second
order in $\alpha$ to calculate all terms of the time dynamics up to $O(\alpha)$. Therefore, it is of
no surprise that the time-dependent terms involving $\Gamma^{(2)}$ can be related to corresponding terms
of the stationary state.

We now discuss our central result (\ref{eq:F_x_0_full}-\ref{eq:F_z_s_full}) and compare it with 
the literature. The leading order term is consistent with the Bloch-Redfield solution 
(\ref{eq:F_0_x}-\ref{eq:F_s_z}), provided one neglects the renormalization of the tunneling. 
Our result shows that the renormalized tunneling appears in a subtle way which can {\it not} be
obtained by just replacing $\Delta\rightarrow\tilde{\Delta}$. There is a $Z$-factor renormalization
$Z=\tilde{\Delta}^2/\Delta^2$ for $F^c_{x,y}$ and $F^s_{x,y}$, and terms $\sim\Delta$ or $\sim\Delta^3$ 
in the Bloch-Redfield solution are replaced by $\sqrt{Z}\Delta=\tilde{\Delta}^2/\Delta$ and 
$Z^2\Delta^3=\tilde{\Delta}^4/\Delta$, respectively.

The most interesting correction in $O(\alpha)$ is the slowly varying logarithmic term 
$\alpha\ln(\Omega t)$ appearing in the function $f_t$ multiplying the leading order terms of 
$F_\alpha^{c/s}$. We note that the correct energy scale in this logarithmic term is the renormalized
Rabi frequency $\Omega$ and {\it not} the Lamb-shift $\Omega-\Omega_0$ as it was 
obtained in Ref.~\onlinecite{divincenzo_loss_05}. As was already mentioned in 
Section~\ref{sec:L_perturbative} via Eq.~(\ref{eq:secular}) the crucial point is not to neglect the 
$O(\alpha)$ contributions in the logarithmic functions. E.g. if one considers the integral 
(\ref{eq:integral_ln}) for $z_i=z_+=\Omega-i\Gamma/2$ and neglects all $O(\alpha)$ 
contributions in the argument of the logarithm by setting 
$\ln(-i(E-z_+)/\Omega)\approx\ln(-i(E-\Omega_0)/\Omega_0)$ one obtains 
\begin{align}
\nonumber
&{i\over 2\pi}\int dE e^{-iEt}{1\over E-z_+}\ln{-i(E-\Omega_0)\over\Omega_0}\,=\\
\label{eq:integral_ln_approx}
& =\,\ln{-i(z_+-\Omega_0)\over\Omega_0}e^{-iz_+ t}
+ H((z_+-\Omega_0)t)e^{-i\Omega_0 t}\quad,
\end{align}
which is obviously quite different from the exact result $-(\gamma + \ln(\Omega t))e^{-iz_+ t}$ 
not only because of the incorrect exponential appearing in the second term on the r.h.s. (which
is just oscillating with the unrenormalized Rabi frequency) but also due to the incorrect
preexponential functions of both terms involving the energy scale of the Lamb shift 
$\delta\Omega=\Omega-\Omega_0$. This shows that the resummation of secular terms
contained in logarithmic contributions of the Liouvillian is not only 
important to get the correct exponential part of the time dynamics but
also to obtain the correct preexponential functions. Only in the limit $\Gamma t,\delta\Omega t\ll 1$,
where $|E-\Omega_0|\gg\delta\Omega,\Gamma$, it is allowed to neglect secular terms by disregarding
the $O(\alpha)$ terms in the argument of the logarithm. In this case one can use the approximation 
$H((z_+-\Omega_0)t)\approx -\gamma-\ln(-i(z_+-\Omega_0)t)$ and 
$e^{-iz_+t}\approx e^{-i\Omega_0t}$ in (\ref{eq:integral_ln_approx}) leading to
\begin{align}
\nonumber
&{i\over 2\pi}\int dE e^{-iEt}{1\over E-z_+}\ln{-i(E-\Omega_0)\over\Omega_0}\,\approx\\
\label{eq:integral_ln_approx_2}
& =\,-\Big(\gamma + \ln(\Omega_0 t)\Big)e^{-i\Omega_0 t}\quad,
\end{align}
with the correct logarithmic time dependence involving the Rabi frequency and {\it not} the
Lamb shift.

For large times $\Omega t\sim 1/\alpha \gg 1$, where the damping is still moderate due to 
$\Gamma t\sim O(1)$, the logarithmic term $\sim\alpha\ln(\Omega t)$ is
the most important correction to the leading order terms. In this regime, the functions 
$H_t$ and $\tilde{H}_t$ lead only to very small contributions and fall off according to
\begin{align}
\label{eq:H_real_large}
H^\prime_t\,&=\,{1\over (\Omega t)^2}\,+\,O\left({1\over(\Omega t)^4}\right)\quad,\\
\label{eq:H_imag_large}
H^{\prime\prime}_t\,&=\,{1\over \Omega t}\,+\,O\left({1\over(\Omega t)^3}\right)\quad,\\
\label{eq:H_tilde_real_large}
\tilde{H}^\prime_t\,&=\,-{1\over (\Omega t)^2}\,+\,O\left({1\over(\Omega t)^4}\right)\quad,\\
\label{eq:H_tilde_imag_large}
\tilde{H}^{\prime\prime}_t\,&=\,O\left({1\over(\Omega t)^3}\right)\quad.
\end{align}
The pure branch cut contributions arising from $H_t$ and $\tilde{H}_t$ are the only terms 
showing a significant time dependence whereas the logarithmic terms are slowly varying in time. The most 
important term is the one arising from $H^{\prime\prime}_t$ which falls off only $\sim 1/(\Omega t)$. It arises
only in the finite bias case for the modes $F^c_{x/z}$ and $F^s_{x/z}$ and has never been reported before.
The standard case treated in the literature \cite{leggett_87,weiss_12} 
is the calculation for the time dynamics of the Pauli matrix in $z$-direction at zero bias 
for the initial condition ${\langle\sigma_{z}\rangle}_0=1$ and ${\langle\sigma_{x/y}\rangle}_0=0$. 
In this case and for $\Omega t\gg 1$ our solution reduces to
\begin{align}
\label{eq:sigma_z_large_times}
{\langle\sigma_{z}\rangle}(t)\,\approx\,
(1+\alpha)\cos(\tilde{\Delta}t)e^{-{\Gamma\over 2}t} 
- 2\alpha{1\over (\tilde{\Delta} t)^2}e^{-\Gamma t}\quad,
\end{align}
which, up to the missing exponential for the second term on the r.h.s., agrees with the 
NIBA result \cite{leggett_87,weiss_12} and the result obtained from the Born approximation
\cite{divincenzo_loss_05} (where also the residuum has been calculated for the first term on
the r.h.s.). In Refs.~\onlinecite{slutskin_etal_11,kashuba_schoeller_13} the correct exponential 
has been obtained for the second term. The important new result for finite bias is that, besides the 
appearance of many other terms falling off $\sim \alpha/(\Omega t)^2$, there are new terms falling
off $\sim \alpha/(\Omega t)$. For $\Omega t\sim 1/\alpha$, this are terms in $O(\alpha^2)$ and 
thus of the same order as other constant terms $\sim\alpha^2$ or slowly varying logarithmic terms
$\sim\alpha^2\ln^2(\Omega t)$ not covered by our analytic solution in the non-exponential regime. 
However, the terms $\sim\alpha/(\Omega t)$ 
are consistent in the sense that they determine the leading behavior of those contributions which
show a significant time dependence. In contrast, terms $\sim\alpha/(\Omega t)^2$ are inconsistent 
in this sense, since for finite bias there will be other strongly varying terms $\sim\alpha^2/(\Omega t)$
of the same order which we have not calculated. Keeping only the consistent terms falling off 
$\sim\alpha/(\Omega t)$ we obtain for {\underline{large times $\Omega t\gg 1$}}:
\begin{widetext}
\begin{align}
\label{eq:F_x_0_large_t}
F_x^0(t) \,&=\,
-{\langle\sigma_x\rangle}_{\text{st}} 
\,-\, \Big(1+2\alpha{\tilde{\Delta}^2\over\Omega^2}\Big){\tilde{\Delta}^2\over\Delta\Omega}
{\langle\tilde{\sigma}_z\rangle}_0
\,+\, \pi\alpha{\tilde{\Delta}^4\epsilon\over\Delta^2\Omega^3}{\langle\sigma_y\rangle}_0
\quad,\\
\label{eq:F_y_0_large_t}
F_y^0(t) \,&=\,
\pi\alpha{\tilde{\Delta}^2\epsilon\over\Delta\Omega^2}(1+{\langle\tilde{\sigma}_z\rangle}_0)
\quad,\\
\label{eq:F_z_0_large_t}
F_z^0(t) \,&=\,
-{\langle\sigma_z\rangle}_{\text{st}}
\,+\, \Big(1+2\alpha{\tilde{\Delta}^2\over\Omega^2}\Big){\epsilon\over\Omega}{\langle\tilde{\sigma}_z\rangle}_0
\,-\, \pi\alpha{\tilde{\Delta}^2\epsilon^2\over\Delta\Omega^3}{\langle\sigma_y\rangle}_0
\quad,\\
\label{eq:F_x_c_large_t}
F_x^c(t) \,&=\,
-f_t{\tilde{\Delta}^2\epsilon\over\Delta^2\Omega}{\langle\tilde{\sigma}_x\rangle}_0
\,+\, {\pi\over 2}\alpha{\tilde{\Delta}^4\epsilon\over\Delta^2\Omega^3}{\langle\sigma_y\rangle}_0
\,+\,2\alpha{\tilde{\Delta}^4\epsilon\over\Delta^2\Omega^3}{1\over \Omega t}{\langle\sigma_y\rangle}_0
\quad,\\
\label{eq:F_y_c_large_t}
F_y^c(t) \,&=\,
f_t{\tilde{\Delta}^2\over\Delta^2}{\langle\sigma_y\rangle}_0
\,-\, \pi\alpha{\tilde{\Delta}^2\epsilon\over\Delta\Omega^2}
\,-\, {\pi\over 2}\alpha{\tilde{\Delta}^4\over\Delta^2\Omega^2}{\langle\tilde{\sigma}_x\rangle}_0
\quad,\\
\label{eq:F_z_c_large_t}
F_z^c(t) \,&=\,
-f_t{\tilde{\Delta}^2\over\Delta\Omega}{\langle\tilde{\sigma}_x\rangle}_0
\,-\,{\pi\over 2}\alpha{\tilde{\Delta}^4\over\Delta\Omega^3}{\langle\sigma_y\rangle}_0
\,+\, 2\alpha{\tilde{\Delta}^2\epsilon^2\over\Delta\Omega^3}{1\over \Omega t}{\langle\sigma_y\rangle}_0
\quad,\\
\label{eq:F_x_s_large_t}
F_x^s(t) \,&=\,
-f_t{\tilde{\Delta}^2\epsilon\over\Delta^2\Omega}{\langle\sigma_y\rangle}_0
\,+\, \pi\alpha{\tilde{\Delta}^2\epsilon^2\over\Delta\Omega^3}
\,+\, 2\alpha{\tilde{\Delta}^4\epsilon\over\Delta^2\Omega^3}
{1\over \Omega t}{\langle\tilde{\sigma}_x\rangle}_0
\,+\, 2\alpha{\tilde{\Delta}^2\epsilon^2\over\Delta\Omega^3}{1\over \Omega t}{\langle\tilde{\sigma}_z\rangle}_0
\quad,\\
\label{eq:F_y_s_large_t}
F_y^s(t) \,&=\,
-f_t{\tilde{\Delta}^2\over\Delta^2}{\langle\tilde{\sigma}_x\rangle}_0
\,-\, \pi\alpha{\tilde{\Delta}^4\over\Delta^2\Omega^2}{\langle\sigma_y\rangle}_0
\quad,\\
\label{eq:F_z_s_large_t}
F_z^s(t) \,&=\,
-f_t{\tilde{\Delta}^2\over\Delta\Omega}{\langle\sigma_y\rangle}_0
\,+\, \pi\alpha{\tilde{\Delta}^2\epsilon\over\Omega^3}
\,+\, 2\alpha{\tilde{\Delta}^2\epsilon\over\Omega^3}{1\over \Omega t}
{\langle\tilde{\sigma}_z\rangle}_0
\,-\, 2\alpha{\tilde{\Delta}^2\epsilon^2\over\Delta\Omega^3}{1\over \Omega t}
{\langle\tilde{\sigma}_x\rangle}_0
\quad.
\end{align}
\end{widetext}

For {\underline{zero bias $\epsilon=0$}} and {\underline{large times $\Omega t\gg 1$}}, we keep the
leading terms falling off $\sim\alpha/(\Omega t)^2$ and obtain with the help of $f_t=1+\alpha$, 
${\langle\sigma_x\rangle}_{\text{st}}=(1+\alpha){\tilde{\Delta}\over\Delta}$,
${\langle\sigma_y\rangle}_{\text{st}}={\langle\sigma_z\rangle}_{\text{st}}=0$, 
${\langle\tilde{\sigma}_x\rangle}_0=-{\Delta\over\tilde{\Delta}}{\langle\sigma_z\rangle}_0$ and 
${\langle\tilde{\sigma}_z\rangle}_0=-{\tilde{\Delta}\over\Delta}{\langle\sigma_x\rangle}_0$ the result
\begin{widetext}
\begin{align}
\label{eq:F_0_zero_bias_large_t}
F_x^0(t) \,&=\,-(1+\alpha){\tilde{\Delta}\over\Delta}
\,+\, (1+2\alpha){\tilde{\Delta}^2\over\Delta^2}{\langle\sigma_x\rangle}_0 \quad,\quad
%
F_y^0(t) \,=\,0 \quad,\quad
%
F_z^0(t) \,=\,-2\alpha{1\over (\tilde{\Delta} t)^2}{\langle\sigma_z\rangle}_0
\quad,\\
\label{eq:F_c_zero_bias_large_t}
F_x^c(t) \,&=\,-2\alpha{1\over (\Delta t)^2}{\langle\sigma_x\rangle}_0
\quad,\quad
%
F_y^c(t) \,=\,
(1+\alpha){\tilde{\Delta}^2\over\Delta^2}{\langle\sigma_y\rangle}_0
\,+\, {\pi\over 2}\alpha{\tilde{\Delta}\over\Delta}{\langle\sigma_z\rangle}_0
\quad,\quad
F_z^c(t) \,=\,
(1+\alpha){\langle\sigma_z\rangle}_0
\,-\,{\pi\over 2}\alpha{\tilde{\Delta}\over\Delta}{\langle\sigma_y\rangle}_0
\quad,\\
\label{eq:F_s_zero_bias_large_t}
F_x^s(t) \,&=\,0 \quad,\quad
F_y^s(t) \,=\,
(1+\alpha){\Delta\over\tilde{\Delta}}{\langle\sigma_z\rangle}_0
\,-\, \pi\alpha{\tilde{\Delta}^2\over\Delta^2}{\langle\sigma_y\rangle}_0
\quad,\quad
F_z^s(t) \,=\,
-(1+\alpha){\tilde{\Delta}\over\Delta}{\langle\sigma_y\rangle}_0
\quad,
\end{align}
\end{widetext}
which agrees with the result obtained in Ref.~\onlinecite{kashuba_schoeller_13}, except that we
have also calculated all time-independent corrections for the preexponential functions in $O(\alpha)$ here. 

For very large times in the 
exponential region where $\alpha\ln(\Omega t)\sim O(1)$, our result is no longer valid and the
RG treatment presented in Section~\ref{sec:exp_large_times} is needed to sum
up all powers of such logarithmic terms (determining the power law exponent in 
leading order in $\alpha$). As we discuss later on, the main result is that the 
function $f_t$ has to be replaced by the power-law
\begin{align}
\label{eq:f_t_exp_region}
f_t\,\rightarrow\,\left({1\over\Omega t}\right)^{2\alpha{\epsilon^2\over\Omega^2}}
\Big(1-2\alpha\gamma{\epsilon^2\over\Omega^2}+\alpha{\tilde{\Delta}^2\over\Omega^2}\Big)\quad.
\end{align}
with a power-law exponent depending on the bias. This power-law exponent is consistent with the one predicted 
in Ref.~\onlinecite{slutskin_etal_11} where $\Omega$ was replaced by the unrenormalized 
Rabi frequency $\Omega_0$. However, in this reference, many terms in higher order in
$\Delta/\Omega_0$ and $\alpha$ have been neglected and a consistent
RG analysis was lacking whether additional logarithmic terms appear which can
change the power-law exponent e.g. from $2\alpha\epsilon^2/\Omega^2$ to $2\alpha$. It turns out
that this analysis depends crucially on the time regime under consideration. Whereas, for exponentially
large times, it turns out that the power-law exponent is indeed $2\alpha\epsilon^2/\Omega^2$ for the
oscillating modes, a completely different result appears for exponentially small times with
a power-law exponent given by $2\alpha$, see (\ref{eq:sigma_x_exp_small_times}) and 
(\ref{eq:sigma_y_exp_small_times}). There is a complicated crossover between these two power laws
since the real part of the functions $H_t^\prime$ and $\tilde{H}_t^\prime$ contain additional 
logarithmic terms for small times $\Omega t\ll 1$, see Eqs.~(\ref{eq:H_real_small}) and 
(\ref{eq:H_tilde_real_small}) below. Only via our consistent RG treatment presented in 
Section~\ref{sec:RTRG} one can be sure to include all terms of the leading logarithmic series
providing the correct power-law exponents in $O(\alpha)$ for exponentially small and large times,
together with the correct crossover behavior in the non-exponential regime. 

One can check that in the limit of small but not exponentially small times our solution 
(\ref{eq:F_x_0_full}-\ref{eq:F_z_s_full}) is consistent with 
(\ref{eq:sigma_x_exp_small_times}-\ref{eq:sigma_z_exp_small_times}). The logarithmic terms
are a result of a combination of logarithmic terms arising from the terms 
$\sim\alpha\ln(\Omega t)$ appearing explicitly in (\ref{eq:F_x_c_full}-\ref{eq:F_z_s_full}) 
and those arising from the functions $H_t$ and $\tilde{H}_t$, which, for small argument, 
can be expanded as
\begin{align}
\label{eq:H_real_small}
H^\prime_t\,&=\,-\gamma\,-\,\ln(\Omega t)\,+\,O(\Omega t)\quad,\\
\label{eq:H_imag_small}
H^{\prime\prime}_t\,&=\,{\pi\over 2}\,+\,O(\Omega t)\quad,\\
\label{eq:H_tilde_real_small}
\tilde{H}^\prime_t\,&=\,-\gamma\,-\,\ln(\Omega t)\,-\,1\,+\,O(\Omega t)\quad,\\
\label{eq:H_tilde_imag_small}
\tilde{H}^{\prime\prime}_t\,&=\,{\pi\over 2}\,+\,O(\Omega t)\quad.
\end{align}
Inserting this expansion in (\ref{eq:F_x_0_full}-\ref{eq:F_z_s_full}) and neglecting all terms
$\sim\alpha\Omega t$ (with or without a logarithm) we obtain
\begin{align}
\nonumber
&{\langle\sigma_{x/y}\rangle}_{\text{st}} + F^0_{x/y} + F^c_{x/y} \,\approx\,\\
\label{eq:xy_small_times}
&\hspace{1cm}
\approx\,{\tilde{\Delta}^2\over\Delta^2}\Big\{1-2\alpha(\gamma + \ln(\Omega t))\Big\}
{\langle\sigma_{x/y}\rangle}_0 \quad,\\
\label{eq:z_small_times}
&{\langle\sigma_z\rangle}_{\text{st}} + F^0_z + F^c_z \,\approx\,{\langle\sigma_z\rangle}_0 \quad,\\
\label{eq:F_s_x_small_times}
&F^s_x \Omega t \,\approx\, -{\tilde{\Delta}^2\over\Delta^2}\epsilon t
{\langle\sigma_y\rangle}_0 \quad,\\
\label{eq:F_s_y_small_times}
&F^s_y \Omega t \,\approx\, {\tilde{\Delta}^2\over\Delta^2}
\Big\{\epsilon t {\langle\sigma_x\rangle}_0 + \Delta t {\langle\sigma_z\rangle}_0 \Big\}\quad,\\
\label{eq:F_s_z_small_times}
&F^s_z \Omega t \,\approx\, -{\tilde{\Delta}^2\over\Delta^2}
\Delta t {\langle\sigma_y\rangle}_0\quad.
\end{align}
Inserting this result in (\ref{eq:time_dynamics_pauli_general_form}), expanding the 
exponential functions up to linear order in $\Omega t$ and again neglecting all terms
$\sim\alpha\Omega t$, we obtain precisely the 
expansion (\ref{eq:sigma_x_not_exp_small_times}-\ref{eq:sigma_z_not_exp_small_times}) for
small but not exponentially small times, showing that we cover the correct crossover behavior by combining the 
solutions (\ref{eq:sigma_x_small_times}-\ref{eq:sigma_z_small_times}) for small or exponentially
small times with (\ref{eq:F_x_0_full}-\ref{eq:F_z_s_full}) in the non-exponential regime. 

For moderate times $\Omega t\sim O(1)$ the logarithmic terms are of the same order as all other
corrections in $O(\alpha)$. In this regime our full solution (\ref{eq:F_x_0_full}-\ref{eq:F_z_s_full})
is needed to calculate all terms one order beyond Bloch-Redfield. In this case the 
time dependence of the preexponential functions is governed by a complicated combination of slowly 
varying logarithmic terms and terms arising from the functions $H_t$ and $\tilde{H}_t$ containing
the exponential integral via (\ref{eq:H_function}) and (\ref{eq:tilde_H_function}). 

Finally, we note that our solution in the non-exponential regime at zero tunneling $\Delta=0$ is 
fully consistent with the exact solution at zero tunneling presented in 
(\ref{eq:sigma_x_zero_tunneling}-\ref{eq:sigma_z_zero_tunneling}) 
and (\ref{eq:zero_tunneling_power_law}). It is even fully reproduced if we use the 
replacement (\ref{eq:f_t_exp_region}) for the function $f_t$. \\

\section{Real-time renormalization group}
\label{sec:RTRG}

In this section we will present the real-time renormalization group approach to calculate
the Liouvillian $L(E)$ beyond perturbation theory by including the leading logarithmic series at
low and high energies. This provides the basis for the renormalized perturbation theory in the
non-exponential regime together with the calculation of the time dynamics for exponentially small or 
large times, see Sections~\ref{sec:L_renormalized_perturbative} and \ref{sec:perturbative}.

\subsection{RG equations}
\label{sec:RG_equations}

The leading order RG equations to determine the Liouvillian $L(E)$ for the ohmic spin boson model 
have been derived in Ref.~\onlinecite{kashuba_schoeller_13}. Using the definitions (\ref{eq:L_sa}), 
(\ref{eq:tilde_L}) and (\ref{eq:L_decomposition}), they read
\begin{widetext}
\begin{align}
\label{eq:rg_tilde_L}
{d\over dE}\tilde{L}_\Delta(E)\,&=\,2\alpha \sum_i Z'(E) G(E) P_i(E) Z'(E) G(E) 
{\tilde{L}_\Delta(E) - \lambda_i(E) \over E - \lambda_i(E)} \quad,\\
\label{eq:rg_Z}
{d\over dE}Z'(E)\,&=\,2\alpha \sum_i Z'(E) G(E) P_i(E) Z'(E) G(E) Z'(E)
{1 \over E - \lambda_i(E)} \quad,\\
\label{eq:rg_G}
{d\over dE}G(E)\,&=\,2\alpha \sum_{ij} G(E) P_i(E) Z'(E) G(E) P_j(E) Z'(E) G(E) 
{{\cal{L}}_i(E) - {\cal{L}}_j(E)  \over \lambda_i(E) - \lambda_j(E)} \quad,
\end{align}
\end{widetext}
with ${\cal{L}}_i(E)=\ln((-i)(E-\lambda_i(E))/\Omega)$, see (\ref{eq:cal_L_Omega}). For $i=j$, we get
$({\cal{L}}_i(E) - {\cal{L}}_j(E))/(\lambda_i(E) - \lambda_j(E)) = -1/(E-\lambda_i(E))$ for 
the last factor on the r.h.s. of (\ref{eq:rg_G}).
Here, $\lambda_i(E)$ and $P_i(E)$ are the eigenvalues and projectors of $\tilde{L}_\Delta(E)$,
respectively, defined in (\ref{eq:right_eigenstate}-\ref{eq:projector}). 
The RG equations for $\tilde{L}_\Delta(E)$ and $Z'(E)$ are coupled to the RG equation for the 
vertex $G(E)$, which is also a $4\times 4$-matrix. They can be solved along a certain path 
in the complex plane with the following initial condition at $E=iD$
\begin{align}
\label{eq:initial_L}
\tilde{L}_\Delta(iD)\,&=\,L_0 \quad,\\
\label{eq:initial_Z}
Z'(iD)\,&=\,\mathbbm{1} \quad,\\
\label{eq:initial_G}
G(iD)\,&=\,
\left(\begin{array}{cc} 0 & 0 \\ 0 & \sigma_z \end{array}\right)\quad.
\end{align} 
In the limit of large $D$, this initial condition can also be taken for $E=\omega-iD$, with 
some real $\omega\ll D$. Fixing the real parameter $\omega$, the RG equations are solved numerically 
along the path $E=\omega-i\Lambda$, starting at $\Lambda=D$. For $\omega=0$ and $\Lambda\rightarrow 0$, 
we obtain the Liouvillian $L(0^+)$ from which we can obtain the stationary state $\rho_{\text{st}}$ via 
(\ref{eq:stationary}). For $\omega=\pm\eta$ ($\eta=0^+$) and $\Lambda < -\Gamma$, we obtain a jump
between $L(-i\Lambda+\eta)$ and $L(-i\Lambda-\eta)$ indicating the branch cut starting at
$z_0=-i\Gamma$. Similarly, for $\omega=\sigma\Omega\pm\eta$ and $\Lambda < -\Gamma/2$, there will
be jump between the two solutions $L(\sigma\Omega-i\Lambda+\eta)$ and $L(\sigma\Omega-i\Lambda-\eta)$,
corresponding to the two branch cuts starting at $z_\sigma=\sigma\Omega-i\Gamma/2$. In this way one
can determine numerically the positions of the branching points $z_i$ ($i=0,\pm$) of the Liouvillian, 
together with fixing the branch cuts along the direction of the negative imaginary axis. This is an 
important advantage of the real-time RG method since it uses the complex Fourier variable $E$ as flow
parameter and, via a numerical solution of the RG equations along a certain path in the complex plane, 
allows for an elegant analytical continuation of retarded quantities into the lower half of the
complex plane. Once the RG equations have been solved in this way the time dynamics 
can be calculated from (\ref{eq:decomposition_time_dynamics}) and (\ref{eq:rho_i_branchcut_integral}) 
with the preexponential operator given in analogy to (\ref{eq:preexponential_operator}) by 
\begin{align}
\nonumber
F_i(t)\,&=\,{1\over 2\pi}\int_0^\infty \,dx \,e^{-xt}\cdot\\
\nonumber
&\hspace{-0.5cm}
\cdot\left\{{1\over E-\tilde{L}_\Delta(E)}Z'(E)\Big|_{E=z_i-ix+\eta}\,-\,(\eta\rightarrow -\eta)\right\}\cdot\\
\label{eq:rg_preexponential_operator}
&\hspace{1cm}
\cdot (1+\Sigma_s{1\over z_i-ix})\rho_0\quad.
\end{align}
Due to the exponential part $e^{-xt}$ in the integrand, the calculation of the integrals is numerically 
very stable, which is the basic reason why the direction of the branch cuts is chosen along 
the negative imaginary axis.

As explained in detail in Refs.~\onlinecite{kashuba_schoeller_13,schoeller_14}, the RG equations
(\ref{eq:rg_tilde_L}-\ref{eq:rg_G}) contain the leading logarithmic series, i.e. power-law exponents
can be calculated reliably up to $O(\alpha)$. All other terms on the r.h.s. of the RG equations are
of higher order in $\alpha$ but scale as function of $E$ in the same way as the leading term, i.e. for
large energies as $1/E$ and for small energies at most as $1/(E-\lambda_i(E))$. Furthermore, the r.h.s. of
the RG equations is universal, i.e. well-defined in the limit $D\rightarrow\infty$, the band width $D$ of 
the bath enters only via the initial value $E=iD$. Therefore, the RG equations provide a well-defined set
of differential equations which can be systematically truncated at order $\alpha$, including secular terms 
and the leading logarithmic series to all orders.  

We note that the vertex renormalization is very essential. At large energies we show in 
Section~\ref{sec:RG_large_E} that the vertex obtains an important $Z$-factor renormalization, which 
is important to determine the correct time dynamics at exponentially small and intermediate times in the
non-exponential regime. For energies close to one of the branching points $z_i$, $i=\text{st},\pm$, the
vertex renormalization is very different, leading to completely different power-laws for the 
preexponential functions of the time dynamics compared to the one at exponentially small times. This
has also been discussed in Ref.~\onlinecite{kashuba_schoeller_13} for the zero-bias case, where it was
shown that $G(E)$ does no longer renormalize for $E$ close to $z_i$. At finite bias, this is quite 
different and leads to the power-law (\ref{eq:f_t_exp_region}), as will be demonstrated in 
Section~\ref{sec:exp_large_times} by a numerical solution of the RG equations. 

Furthermore, we note that the vertex renormalization has to be taken with care when solving the
RG equations along a certain branch cut. For $i\ne j$, the difference ${\cal{L}}_i(E)-{\cal{L}}_j(E)$ 
can contain a constant imaginary term due to the jump of the various logarithm across the branch cuts. Although
this constant term does not lead to a logarithmic contribution and contributes inconsistent terms in
$O(\alpha^2)$ for the Liouvillian, it leads to a term $\sim \alpha |E-z_i|/\Omega$ for the vertex, 
which becomes of $O(1)$ for $|E-z_i|\sim\Omega/\alpha$. This effect is inconsistent and is canceled 
by other contributions from higher orders. However, this effect can easily be avoided by just omitting
this constant imaginary term for the vertex renormalization when $E=z_i-ix\pm\eta$ with $x>\Omega$. 
This is consistent with the analytic solution of the RG equations up to $O(\alpha)$ for large energies and 
energies in the non-exponential regime, as shown in Sections~\ref{sec:RG_large_E} and \ref{sec:RG_non_exp_E}.

\subsection{Large energies}
\label{sec:RG_large_E}

For large energies, where $\Omega \ll |E| \ll D$, we will show in this section that $\tilde{L}_\Delta(E)$ and
$Z'(E)$ are given by (\ref{eq:tilde_L_delta_large_energies}-\ref{eq:Z_large_E}). For $|E|\gg|\lambda_i(E)|$,
the RG equations (\ref{eq:rg_tilde_L}-\ref{eq:rg_G}) can be approximated by
\begin{widetext}
\begin{align}
\nonumber
{d\over dE}\tilde{L}_\Delta(E)\,&=\,2\alpha \sum_i Z'(E) G(E) P_i(E) Z'(E) G(E) 
{\tilde{L}_\Delta(E) - \lambda_i(E) \over E}\\ 
\label{eq:rg_tilde_L_large_E}
\,&=\,2\alpha\Big\{Z'(E) G(E) Z'(E) G(E) \tilde{L}_\Delta(E) -  Z'(E) G(E) \tilde{L}_\Delta(E) Z'(E) G(E)\Big\}
{1\over E} \quad,\\
\label{eq:rg_Z_large_E}
{d\over dE}Z'(E)\,&=\,2\alpha Z'(E) G(E) Z'(E) G(E) Z'(E) {1 \over E } \quad,\\
\label{eq:rg_G_large_E}
{d\over dE}G(E)\,&=\,-2\alpha G(E) Z'(E) G(E) Z'(E) G(E) {1 \over E} \quad.
\end{align}
\end{widetext}
With the initial conditions (\ref{eq:initial_L}-\ref{eq:initial_G}) these RG equations can be solved 
by the ansatz 
\begin{align}
\label{eq:tilde_L_ansatz_large_E}
\tilde{L}_\Delta(E)\,&=\,
\left(\begin{array}{cc} 0 & \Delta\tau_- \\ \Delta Z(E) \tau_- & \epsilon \sigma_z \end{array}\right)
\quad,\\
\label{eq:Z_ansatz_large_E}
Z'(E)\,&=\,
\left(\begin{array}{cc} 1 & 0 \\ 0 & Z(E) \end{array}\right)
\quad,\\
\label{eq:G_ansatz_large_E}
G(E)\,&=\,g(E)
\left(\begin{array}{cc} 0 & 0 \\ 0 & \sigma_z \end{array}\right)
\quad.
\end{align}
Inserting this ansatz into the r.h.s. of the RG equations we find
\begin{align}
\label{eq:rg_tilde_L_ansatz_large_E}
{d\over dE} \tilde{L}_\Delta(E)\,&=\,2\alpha Z(E)^2 g(E)^2 {1\over E}
\left(\begin{array}{cc} 0 & 0 \\ \Delta Z(E) \tau_- & 0 \end{array}\right)
\quad,\\
\label{eq:rg_Z(E)_large_E}
{d\over dE}Z(E)\,&=\,2\alpha Z(E)^3 g(E)^2 {1\over E}\quad,\\
\label{eq:rg_g(E)_large_E}
{d\over dE}g(E)\,&=\,-2\alpha Z(E)^2 g(E)^3 {1\over E}\quad,
\end{align}
showing that (\ref{eq:rg_tilde_L_ansatz_large_E}) is consistent with the ansatz 
(\ref{eq:tilde_L_ansatz_large_E}). Furthermore we find
\begin{align}
\label{eq:rg_Zg_large_E}
{d\over dE} Z(E)^2 g(E)^2\,&=\,0\quad,\\
\label{eq:rg_Z(E)_2_large_E}
{d\over dE} Z(E)\,&=\,2\alpha Z(E) {1\over E}\quad,
\end{align}
with the solution
\begin{align}
\label{eq:Zg_large_E}
Z(E)^2 g(E)^2\,&=\, 1\quad,\\
\label{eq:Z(E)_large_E}
Z(E)\,&=\,\left({-iE\over D}\right)^{2\alpha}\quad.
\end{align}
In conclusion, (\ref{eq:tilde_L_ansatz_large_E}), (\ref{eq:Z_ansatz_large_E}) and 
(\ref{eq:Z(E)_large_E}) prove the form (\ref{eq:tilde_L_delta_large_energies}-\ref{eq:Z_large_E}) of 
the Liouvillian at large energies which was used in Section~\ref{sec:small_times} to calculate the
time dynamics for small times.

\subsection{The non-exponential regime}
\label{sec:RG_non_exp_E}

In the non-exponential regime (\ref{eq:non_exp_E}), where $\alpha\ln(-i(E-\lambda_i(E))/\Omega)\ll 1$,
the RG equations can be solved perturbatively around the solution 
(\ref{eq:tilde_L_ansatz_large_E}-\ref{eq:G_ansatz_large_E}) at high energies evaluated at 
$E=i\Omega$, see Ref.~\onlinecite{schoeller_09} for details. Denoting the latter by
\begin{align}
\label{eq:tilde_L_Omega_large_E}
\tilde{L}_0\,&=\,
\left(\begin{array}{cc} 0 & \Delta\tau_- \\ \Delta Z \tau_- & \epsilon \sigma_z \end{array}\right)
\quad,\\
\label{eq:Z_Omega_large_E}
Z^\prime\,&=\,
\left(\begin{array}{cc} 1 & 0 \\ 0 & Z \end{array}\right)
\quad,\quad Z\,=\,\left({\Omega\over D}\right)^{2\alpha}\,=\,{\tilde{\Delta}^2\over\Delta^2}\quad,\\
\label{eq:G_Omega_large_E}
G\,&=\,g
\left(\begin{array}{cc} 0 & 0 \\ 0 & \sigma_z \end{array}\right)\quad,\quad g\,=\,1/Z
\quad,
\end{align}
the solution of the RG equations in the non-exponential regime can be written up to $O(\alpha)$ as
\begin{align}
\nonumber 
\tilde{L}_\Delta(E)\,&=\,\tilde{L}_0 \,+\\
\label{eq:tilde_L_nonexp}
&\hspace{-1cm}
 +\,2\alpha \sum_i Z' G P_i^{(0)} Z' G 
(\tilde{L}_0-\lambda_i(E)){\cal{L}}_i(E)\quad,\\
\label{eq:Z_nonexp}
Z'(E)\,&=\,Z' \,+\,2\alpha \sum_i Z' G P_i^{(0)} Z' G Z'{\cal{L}}_i(E)\quad,
\end{align}
with ${\cal{L}}_i(E)=\ln(-i(E-\lambda_i(E))/\Omega)$ defined in
(\ref{eq:cal_L}). Our convention for the notation of $Z'$ and $G$ is chosen
such that when no argument $E$ is written, we implicitly take the high-energy solution 
evaluated at $E=i\Omega$, given by (\ref{eq:Z_Omega_large_E}-\ref{eq:G_Omega_large_E}). 
$P_i^{(0)}$ denote the projectors of $\tilde{L}_0$ in lowest order in $\alpha$, 
which are given by (\ref{eq:P_0_i_0}-\ref{eq:P_0_i_sigma})
\begin{align}
\label{eq:P_0_0}
A P^{(0)}_0 A\,&=\,
\left(\begin{array}{cc} \tau_- & 0 \\ 0 & 0  \end{array}\right)\quad,\\
\label{eq:P_0_sigma}
A P^{(0)}_\sigma A\,&=\,{1\over 2}
\left(\begin{array}{cc} 0 & 0 \\ 0 & 1+\sigma\sigma_z \end{array}\right)\quad,
\end{align}
where the matrix $A$ is defined in (\ref{eq:matrix_A}). For large $|E|\gg|\lambda_i(E)|$ but not 
exponentially large the solution (\ref{eq:tilde_L_nonexp}-\ref{eq:Z_nonexp})
is consistent with the result at large energies, given by (\ref{eq:tilde_L_ansatz_large_E}), 
(\ref{eq:Z_ansatz_large_E}) and (\ref{eq:Z(E)_large_E}), when expanded in $\alpha\ln(-iE/\Omega)$.
As a consequence it is straightforward to see that (\ref{eq:tilde_L_nonexp}-\ref{eq:Z_nonexp}) is
indeed the solution of the RG equations in the non-exponential regime up to $O(\alpha)$ since the
differential equation is fulfilled and the boundary condition at large energies is reproduced.

Using (\ref{eq:tilde_L}) and (\ref{eq:L_decomposition}), we find from the solution 
(\ref{eq:tilde_L_nonexp}-\ref{eq:Z_nonexp}) the following result for the Liouvillian $L_a(E)$ up to
$O(\alpha)$
\begin{align}
\label{eq:L_a_nonexp}
L_a(E)\,&=\,E \,-\, {1\over Z'}(E-\tilde{L}_0+\tilde{\Sigma}_a(E))\\
\label{eq:sigma_a_nonexp}
\tilde{\Sigma}_a(E)\,&=\,2\alpha\sum_i Z' G P_i^{(0)} Z' G {\cal{F}}_i(E)\quad,
\end{align}
with ${\cal{F}}_i(E)=(E-\lambda_i(E)){\cal{L}}_i(E)$ defined in (\ref{eq:F}). Inserting the algebra
of the various matrices by using (\ref{eq:Z_Omega_large_E}-\ref{eq:G_Omega_large_E}) and 
(\ref{eq:P_0_0}-\ref{eq:P_0_sigma}), we obtain (\ref{eq:renormalized_propagator}) for the 
propagator together with (\ref{eq:renormalized_sigma_a}) for $\tilde{\Sigma}_a(E)$. 

To check that (\ref{eq:renormalized_lambda_0}-\ref{eq:renormalized_lambda_sigma}) are indeed the
eigenvalues of $\tilde{L}_\Delta(E)$, we transform (\ref{eq:tilde_L_nonexp}) with the matrix $A$ 
and replace $\lambda_i(E)$ by the eigenvalues of $\tilde{L}_0$. After a straightforward calculation 
we obtain
\begin{align}
\nonumber
A\tilde{L}_\Delta(E)A&= 
\Omega\left(\begin{array}{cc} 0 & 0 \\ 0 & \sigma_z \end{array}\right)+
2\alpha{\tilde{\Delta}^2\over\Omega}{\cal{L}}_0(E) 
\left(\begin{array}{cc} 0 & 0 \\ 0 & \tau_-\sigma_z \end{array}\right)\\
\label{eq:tilde_L_nonexp_explicit}
& \hspace{-1cm}
-\alpha{\tilde{\Delta}^2\over\Omega}\sum_\sigma \sigma{\cal{L}}_\sigma(E)
\left(\begin{array}{cc} \tau_- & 0 \\ {\epsilon\over\Delta} (\sigma_z+\sigma)\tau_- & 0 \end{array}\right)\,.
\end{align}
As a consequence the four eigenvalues are given up to $O(\alpha)$ by
\begin{align}
\label{eq:eigenvalue_st_tilde_L}
\lambda_{\text{st}}(E)\,&=\,0\quad,\\
\label{eq:eigenvalue_0_tilde_L}
\lambda_0(E)\,&=\,-\alpha{\tilde{\Delta}^2\over\Omega}\sum_\sigma\sigma{\cal{L}}_\sigma(E)\quad,\\
\label{eq:eigenvalue_sigma_tilde_L}
\lambda_\sigma(E)\,&=\,\sigma(\Omega+\alpha{\tilde{\Delta}^2\over\Omega}{\cal{L}}_0(E))\quad,
\end{align}
in agreement with (\ref{eq:renormalized_lambda_0}-\ref{eq:renormalized_lambda_sigma}).

\subsection{Exponentially large times}
\label{sec:exp_large_times}

For exponentially large times, where higher powers in $\alpha\ln(\Omega t)$ become significant and can
no longer be treated in lowest order to analyze the corrections to Bloch-Redfield, we need a solution 
of the RG equations exponentially close to the branching points $z_i$. Analytically, such an analysis 
is very complicated for arbitrary bias but can be done at zero bias, see Ref.~\onlinecite{kashuba_schoeller_13}.
For arbitrary bias, we have studied the numerical solution of the RG equations and will present a fit to an
analytical ansatz in this section. 

The case of {\underline{zero bias $\epsilon=0$}} has been studied in Ref.~\onlinecite{kashuba_schoeller_13} 
by using the real-time RG method. The main result was that the result 
(\ref{eq:F_0_zero_bias_large_t}-\ref{eq:F_s_zero_bias_large_t}) for large times still holds for 
exponentially large times, except for $F_x^c(t)$, which obtains an additional function $s_0(t)$
\begin{align}
\label{eq:exp_large_times_zero_bias}
F_x^c(t)\,=\,-2\alpha{s_0(t)\over (\Delta t)^2}{\langle\sigma_x\rangle}_0\quad,
\end{align}
with
\begin{align}
\label{eq:s_0}
s_0(t)\,=\,\left({1\over (1+\alpha\ln(\Omega t))(1-\ln(1+\alpha\ln(\Omega t)))}\right)
\quad,
\end{align}
such that the complete solution for ${\langle\sigma_x\rangle}(t)$ reads
\begin{align}
\nonumber
{\langle\sigma_x\rangle}(t)\,&=\,{\langle\sigma_x\rangle}_{\text{st}}(1-e^{-\Gamma t})
\,+\,(1+2\alpha){\tilde{\Delta}^2\over\Delta^2}e^{-\Gamma t}{\langle\sigma_x\rangle}_0\\
\label{eq:sigma_x_zero_bias}
&\hspace{0.5cm}
-\,2\alpha{s_0(t)\over (\Delta t)^2}\cos(\Omega t) e^{-\Gamma t/2}{\langle\sigma_x\rangle}_0\quad,
\end{align}
with ${\langle\sigma_x\rangle}_{\text{st}}=(1+\alpha){\tilde{\Delta}^2\over\Delta^2}$. However,
this result is not very important since, at zero bias, the importance of higher orders in
$\alpha\ln(\Omega t)$ for the preexponential function shows only up for $\Gamma t\gg 1$, where the
exponential damping leads to a negligible result for the time dynamics. Only for $\alpha\sim 1$,
the estimation in (\ref{eq:time_scales_exp}) shows that higher powers of logarithmic terms are
important for times where the damping is moderate. Only from an academic point of view, where the
preexponential function can be studied separately, exponentially large times are also interesting
at zero bias and the function $s_0(t)$ can be identified. As already discussed in 
Ref.~\onlinecite{kashuba_schoeller_13}, we note that there is no change of the power-law exponent of
the $1/t^2$ parts, in particular for the time dynamics of ${\langle\sigma_z\rangle}(t)$, see
(\ref{eq:sigma_z_large_times}), in contrast to the NIBA solution which predicts an incorrect 
power-law exponent $2-2\alpha$ \cite{leggett_87,weiss_12}.

\begin{figure}[htbp!]
  \includegraphics[width=\linewidth]{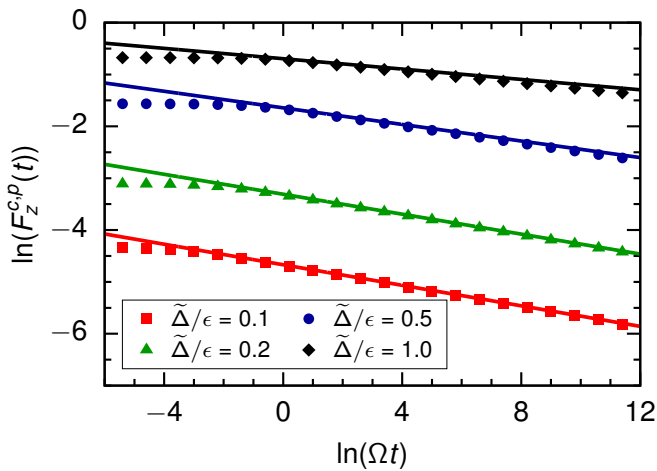}
  \caption{The numerical solution for the logarithm of the pole contribution $F_z^{c,p}(t)$ 
    for $\alpha=0.05$, ${\langle\sigma_z\rangle}_0=1$, 
    ${\langle\sigma_x\rangle}_0={\langle\sigma_y\rangle}_0=0$, and
    different values for $\tilde{\Delta}/\epsilon$ plotted as function of $\ln(\Omega t)$. 
    For very large times a power-law appears with exponent $2\alpha\epsilon^2/\Omega^2$, in
    agreement with (\ref{eq:F_z_c_full_pole}) (solid lines). The band width is chosen as $D/\Delta=10^6$.}
\label{fig:power_law}
\end{figure}

As discussed in detail via the estimation (\ref{eq:time_scales_exp}), the importance of higher orders 
in $\alpha\ln(\Omega t)$ changes significantly for large bias.
At arbitrary bias, we have checked numerically that a power-law appears for the leading-order 
term of the oscillating modes in the regime of very large times, with a bias-dependent exponent 
$2\alpha\epsilon^2/\Omega^2$.  
E.g. Fig.~\ref{fig:power_law} shows the numerical solution for the pole contribution of $F_z^{p,c}(t)$ 
(i.e. the first term on the r.h.s. of (\ref{eq:F_z_c_full})) for various values of the bias. 
For large times, the logarithm of this contribution shows indeed a straight line as function of 
$\ln(\Omega t)$ with a slope given by $-2\alpha\epsilon^2/\Omega^2$
\begin{align}
\label{eq:F_z_c_full_pole}
\ln(F_z^{p,c}(t))\,=\,-2\alpha{\epsilon^2\over\Omega^2}\ln(\Omega t)\,+\,\text{const}\quad,
\end{align}
where the constant term on the r.h.s. is independent of time but depends on the bias.  

\section{Summary}
\label{sec:summary}

In this work we have presented the solution for the time dynamics of the ohmic spin boson 
model at finite bias by systematically expanding one order beyond Boch-Redfield. 
Using real-time RG and perturbation theory we have set up a renormalized 
perturbation theory to study analytically the whole time regime from exponentially small 
($\Omega t\sim e^{-1/\alpha}$) up to large times ($\Omega t\gg 1$). For very large times we used the
real-time RG method to sum up the leading logarithmic series in $\alpha\ln(\Omega t)$. As a result
we obtained several interesting features for the time dynamics: (1) We showed how both the 
unrenormalized ($\Delta$) and renormalized tunneling ($\tilde{\Delta}$) enter the time dynamics 
and that it is not possible to account for the renormalization by using a local Hamiltonian 
with a renormalized tunneling. As in Ref.~\onlinecite{divincenzo_loss_05} we found that the 
renormalized Rabi frequency enters as high-energy cutoff scale to determine $\tilde{\Delta}$.
(2) We found that all terms of the time evolution are exponentially damped by summing up all
secular terms $\sim (\Gamma t)^n$. This results from a self-consistent perturbation theory in analogy
to the one presented in Ref.~\onlinecite{slutskin_etal_11}. (3) For the preexponential functions of
the oscillating modes and in the non-exponential time regime we found logarithmic terms
$\sim\alpha\ln(\Omega t)$ containing the renormalized Rabi frequency as energy scale together with
terms falling off as $\alpha/(\Omega t)$. (4) We showed that some correction terms in $O(\alpha)$ to 
Bloch-Redfield require an analysis of the Liouvillian up to second order in $\alpha$. We were able
to calculate these terms by relating them to the stationary density matrix. (5) By resumming the
leading logarithmic series in $\alpha\ln(\Omega t)$ in all orders of perturbation theory we found
for the preexponential functions of the oscillating modes an interesting crossover from
a power-law $\sim 1/(\Omega t)^{2\alpha}$ at exponentially small times to a power-law   
$\sim 1/(\Omega t)^{2\alpha{\epsilon^2\over\Omega^2}}$ at exponentially large times. The latter has also
been proposed in Ref.~\onlinecite{slutskin_etal_11} but the logarithms determining the crossover to
the power-law at small times have not been discussed there. 

We have identified three important reasons why it is not sufficient to calculate the kernel of
the kinetic equation up to first order in the coupling to the bath to obtain all terms of the first correction 
to the Bloch-Redfield result. We now discuss why these issues are quite generic and are expected to occur
also for other models of dissipative quantum mechanics. 

First, for times of the order of the inverse decay rate $t\sim\Gamma^{-1}$, where damping is still  
moderate, the distance of the Fourier variable $E$ to some of the poles $z_i$ of the
propagator is proportional to the decay rate $|E-z_i|\sim\Gamma$. In this case perturbation theory is 
quite subtle since the denominator $E-L(E)$ of the propagator is of $O(\Gamma)$. The kernel $\Sigma(E)$ 
can no longer be considered as a small
correction compared to $E-L_0$ and can not be expanded up to the numerator. We solved this problem
by expanding all analytic parts of $\Sigma(E)$ around $E=z_i$ and keeping $\Sigma(z_i)$ in the denominator
whereas all other higher terms of the Taylor expansion are at least of $O(\alpha^2)$ and can be taken as
a small correction. The non-analytic terms of $\Sigma(E)$ are more subtle and are some function 
$f_i(E-z_i)$ when $E$ is close to $z_i$, where $f_i(E)~\sim\alpha$ is a non-analytic function 
with branch cut on the negative imaginary 
axis. For the ohmic spin boson model we get $f_i(E-z_i)\sim\alpha (E-z_i)\ln(-i(E-z_i))\sim\alpha^2$ 
such that it can be considered as a small correction. For dissipative quantum models with logarithmic
divergencies at high and low energies it is typical that $\Sigma(E)$ has a logarithmic form,
see e.g. the Kondo model \cite{schoeller_reininghaus_09} or the interacting resonant level model
\cite{RTRG_irlm}, see Ref.~\onlinecite{schoeller_14} for a review. For weak coupling problems and $E$ close
to $z_i$, $\Sigma(E)$ contains either logarithmic terms $\sim\ln(-i(E-z_j))$ with branching points 
$z_j\ne z_i$ (i.e. are analytic and can be expanded around $E=z_i$) 
or are proportional to $(E-z_i)\ln(-i(E-z_i))$ (such that they vanish at
$E=z_i$). Terms $\sim\Gamma\ln(-i(E-z_i))$ with a constant energy scale in front diverge at $E=z_i$ and
are typical for strong coupling problems like e.g. the Kondo model. Most importantly, even for weak coupling
problems, it is never allowed to expand any part of $\Sigma(E)$ in $\alpha$ by setting 
$z_i=z_i^{(0)}+\delta z_i$, where $z_i^{(0)}$ are the pole positions without the bath and 
$\delta z_i\sim O(\alpha)$ denotes the correction from the bath, since $(E-z_i^{(0)})/\delta z_i$ is a parameter 
of $O(1)$. Thus, for any model of dissipative quantum mechanics, it is very dangerous to use a naive 
perturbative expansion of the kernel in the coupling to the bath. The positions $z_i$ of the branching points 
of $L(E)$ (or poles of the propagator) should be kept non-perturbatively in a self-consistent way by
using the full propagator and {\it not} the bare one between the vertices, as also emphasized in
Ref.~\onlinecite{slutskin_etal_11}. For a non-interacting bath described by a quadratic form 
$H_{\text{bath}}=\sum_q\omega_q a_q^\dagger a_q$ in the field operators, the diagrammatic technique 
developed in Ref.~\onlinecite{schoeller_09} shows that all bare propagators can be replaced by
full ones without any double counting such that a systematic self-consistent perturbation theory can be
set up. Whether this is also possible for more complicated baths like e.g. spin baths is an open 
question. 

Secondly, we have seen that degenerate perturbation theory is generically needed since the decay poles 
$z_i=-i\Gamma_i$ and the stationary pole $z_{\text{st}}=0$ of the propagator are close to each other within
the decay rate $\Gamma_i\sim\alpha$. Therefore, second order terms are needed for the Liouvillian 
to calculate the stationary state and all terms of the time evolution of the purely decaying modes
up to first order in $\alpha$. Again this problem occurs only
for times of the order of the inverse decay rate, since for small times $|E-L_0|\sim 1/t$ is 
much larger than $\Gamma$ and can be considered as the largest term in the denominator of the
propagator such that the full kernel $\Sigma(E)$ can be expanded up to the numerator. Thus, for two state
models with one purely decaying and two oscillating modes, the complicated terms in the stationary state and the 
purely decaying mode arising from the second order terms of the Liouvillian, must generically cancel for
small times. This simplifies the calculation of those terms for the purely decaying mode since they
can be expressed via the stationary state which, for the equilibrium case, can be easily calculated 
up to first order in $\alpha$ via the partition function. 
This strategy has been taken over in this work by using the stationary state 
calculated in Ref.~\onlinecite{divincenzo_loss_05} up to $O(\alpha)$. However, for generic models with more than
two local states, several purely decaying modes can occur and the problem of degenerate perturbation theory
can no longer be solved by just calculating the first correction to Bloch-Redfield of the stationary state. 

Whereas the two aforementioned issues are important to be considered for the calculation of the first correction 
to Bloch-Redfield on {\it all} time scales, there are further problems with weak coupling expansions in the
regimes of exponentially small or large times. They arise for problems of dissipative quantum mechanics 
with logarithmic divergencies at high and low energies like the ohmic spin boson model, the interacting
resonant level model, quantum dot models and the Kondo model. They have to be 
treated by an appropriate renormalization group method like the RTRG method \cite{schoeller_09,schoeller_14}. 
For weak coupling problems, where the renormalized vertices stay small in the whole complex plane, the
RG equations can be truncated systematically such that logarithmic terms are summed up
non-perturbatively in leading or subleading order.  
Whereas logarithmic divergencies at high energies can be incorporated in renormalized parameters from poor man
scaling equations, logarithmic divergencies at low energies close to the branching points $z_i$ are quite
subtle and require a full solution of the RG equations. For models of dissipative quantum mechanics without
logarithmic divergencies this issue is not important.

\section*{Acknowledgments}
This work was supported by the Deutsche Forschungsgemeinschaft via RTG 1995. 
We thank D. Loss and D. diVincenzo for valuable discussions.

\begin{appendix}

\section{Liouvillian in perturbation theory}
\label{sec:appendix_L_perturbative}

Here we calculate the Liouvillian up to first order in the coupling $\alpha$ to the bath by using
the diagrammatic technique developped in Ref.~\onlinecite{kashuba_schoeller_13}, where an expansion
in the coupling to the bath is used together with the application of Wick's theorem to integrate out
the phonon bath. In this reference it is shown for the ohmic spin boson model that the 
kernel $\Sigma(E)=\Sigma_s+\Sigma_a(E)$ can be split into two parts,
one stemming from the symmetric and one from the antisymmetric part of the Bose distribution function of the
bath. At zero temperature this leads to Eq.~(\ref{eq:L_sa}) with $\Sigma_s$ given by (\ref{eq:sigma_s}). The 
antisymmetric part $\Sigma_a(E)$ involves only the antisymmetric part of the Bose distribution $n(\omega)$
of the bath 
\begin{align}
\label{eq:antisymmetric_bose}
n_a(\omega)\,=\,{1\over 2}(n(\omega)-n(-\omega)) \,=\,{1\over 2}\text{sign}(\omega)
\end{align}
since $n(\omega)=-\theta(-\omega)$ at zero temperature. The lowest order diagram for
%
\begin{figure}
  \centering
  \includegraphics[width=0.5\hsize]{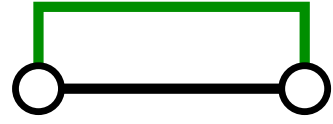}
  \caption{The lowest order diagram for the kernel $\Sigma_a(E)$. Here, the circles 
    represent the bare vertices $G$, the black line connecting the vertices is the 
    local propagator and the green line denotes the bath contraction.}
\label{fig:lowest_order}
\end{figure}
$\Sigma_a(E)$ is shown in Fig.~\ref{fig:lowest_order}, where the green line indicates the contraction 
between the bath field operators which involves the spectral density (\ref{eq:ohmic_spectral_density})
of the bath and the antisymmetric part of the Bose distribution function via
\begin{align}
\label{eq:bath_contraction}
\gamma_a(\omega)\,=\,2\alpha\omega J_c(\omega) n_a(\omega)
\,=\,\alpha|\omega|{D^2\over D^2+\omega^2}\quad.
\end{align}
Using the diagrammatic rules the diagram is translated as
\begin{align}
\label{eq:sigma_a_diagram}
\Sigma_a(E)\,=\,\int d\omega \gamma_a(\omega) G R_a(E+\omega) G \quad,
\end{align}
where $G$ is the bare vertex given by (\ref{eq:initial_G}) and $R_a(E)=1/(E-L_a(E))$ is the local
propagator of the antisymmetric part only. 
To approximate the $\omega$-dependence of $R_a(E+\omega)$, we exhibit the logarithmic
parts by using the decomposition (\ref{eq:resolvent_tilde_L}) and use the spectral decomposition
(\ref{eq:right_eigenstate}-\ref{eq:projector}) of $\tilde{L}_\Delta(E)$
\begin{align}
\nonumber
R(E+\omega)\,&=\,{1\over E+\omega-\tilde{L}_\Delta(E+\omega)}Z'(E+\omega)\\
\label{eq:propagator_decomposition}
&\hspace{-1cm}
=\,\sum_i {1\over E+\omega-\lambda_i(E+\omega)}P_i(E+\omega) Z'(E+\omega)\quad.
\end{align}
Neglecting the $\omega$-dependence of the logarithmic functions $\lambda_i(E+\omega)$, 
$P_i(E+\omega)$ and $Z'(E+\omega)$ (leading to higher orders in $\alpha$), and using the
integral (defined for $\text{Im}(E)>0$ and analytically continued into the lower half of
the complex plane by choosing the branch cut along the direction of the
negative imaginary axis)
\begin{align}
\nonumber
\int d\omega |\omega| {D^2\over D^2 + \omega^2} {1\over E+\omega}
\,&=\,{D^2\over D^2 + E^2} 2E \ln{-iE\over D} \\
\label{eq:integral_resolvent}
&
\xrightarrow{D\rightarrow\infty} \,2E \ln{-iE\over D}\quad,
\end{align}
where $\ln(z)$ is the natural logarithm with branch cut on the negative real axis,
we find from (\ref{eq:sigma_a_diagram})
\begin{align}
\label{eq:sigma_a_2}
\Sigma_a(E)\,=\,2\alpha \sum_i {\cal{F}}_i(E) G P_i(E)Z'(E) G \quad,
\end{align}
with ${\cal{F}}_i(E)$ defined in (\ref{eq:F}-\ref{eq:cal_L}). Taking the
projectors $P_i(E)$ and $Z'(E)$ in lowest order, 
given by (\ref{eq:P_0_lowest_order}-\ref{eq:P_sigma_lowest_order}) and $Z'(E)^{(0)}=1$,
and inserting (\ref{eq:initial_G}) for $G$, we find the result (\ref{eq:sigma_a}-\ref{eq:hat_M_sigma}).

We note that the non-analytic features of $\Sigma_a(E)$ in the lower half of the complex plane are
located at $E=z_i-ix$, $0<x<\infty$, where $z_i$ are the positions of the poles of $R_a(E)$. This holds 
exactly and can be shown in all orders of perturbation theory \cite{schoeller_09,schoeller_14}. E.g. for the 
lowest order diagram (\ref{eq:sigma_a_diagram}) we can see that this holds even when we do not use
any approximation for the $\omega$-dependence of $R_a(E+\omega)$. Closing the integration contour
in the upper half and noting that $R_a(E+\omega)$ is an analytic function there and $\gamma_a(E)$ has
non-analytic features only on the imaginary axis, we find the result
\begin{align}
\nonumber
\Sigma(E)\,&=\,i\int_0^\infty dx \Big\{\gamma_a(ix+0^+)-\gamma_a(ix-0^+)\Big\}\cdot \\ 
\label{eq:sigma_a_nonanalytic}
& \hspace{2cm}
\cdot G R_a(E+ix) G \quad.
\end{align}
Since $R_a(E+ix)$ has a pole at $E+ix=z_i$, we find that $\Sigma(E)$ is non-analytic for
$E=z_i-ix$ with $0<x<\infty$. A similar proof can be used to show this in all orders of perturbation theory,
see Ref.~\onlinecite{schoeller_09,schoeller_14}. 

Furthermore, we note that the matrix structure
\begin{align}
\label{eq:sigma_a_matrix_structure}
\Sigma_a(E)\,=\,\left(\begin{array}{cc} 0 & 0 \\ 0 & \hat{\Sigma}_a(E) \end{array}\right)
\end{align}
holds in all orders of perturbation theory. This is due to the fact that the bare vertices $G$ have
the same structure, see (\ref{eq:initial_G}), and for each diagram all intermediate propagators are
sandwiched between two vertices. A consequence of this matrix structure is that
the projector $P_\text{st}$ on the zero eigenvalue of $\tilde{L}_\Delta(E)$ is exactly known and given
by (\ref{eq:P_st}).

\section{Projectors for $\tilde{L}_0+\tilde{\Sigma}_a^i$}
\label{sec:appendix_projectors_tilde_L_a}

To calculate the projectors of the matrix $\tilde{L}_0+\tilde{\Sigma}_a^i$ up to $O(\alpha)$, 
we first set up the matrix $\tilde{\Sigma}_a^i=\tilde{\Sigma}_a(z_i)$ by setting $E=z_i$ in  
(\ref{eq:renormalized_sigma_a}) and use
\begin{align}
\nonumber
& {\cal{F}}_0(0)\,\sim\,O(\alpha)\quad,\quad
{\cal{F}}_0(z_0)\,=\,{\cal{F}}_\sigma(z_\sigma)\,=\,0 \quad,\\
\nonumber
& {\cal{F}}_\sigma(0)\,,\,{\cal{F}}_\sigma(z_0)\,,\,{\cal{F}}_0(z_\sigma) 
\,=\,-i{\pi\over 2}\Omega + O(\alpha) \quad,\\
& {\cal{F}}_{-\sigma}(z_\sigma)\,=\,2\sigma\Omega\ln{2}-i\pi\Omega+O(\alpha) \quad,
\end{align}
This gives for $\tilde{\Sigma}_a^i$ transformed with the matrix $A$
(see (\ref{eq:matrix_A})) up to $O(\alpha)$ the result
\begin{align}
\label{eq:sigma_a_0_matrix_structure}
A\tilde{\Sigma}_a^\text{st} A\,&=\,A\tilde{\Sigma}_a^0 A\,=\,-i\pi\alpha{1\over\Omega}
\left(\begin{array}{cc} \tilde{\Delta}^2 \tau_- & \Delta\epsilon\tau_-\sigma_z \\ 
{\tilde{\Delta}^2\over\Delta}\epsilon\sigma_z\tau_-   & \epsilon^2 \end{array}\right) \quad,\\
\nonumber
A\tilde{\Sigma}_a^\sigma A\,&=\,-i\pi\alpha{\tilde{\Delta}^2\over\Omega}
\left(\begin{array}{cc} 0 & 0 \\ 0   & \tau_- \end{array}\right) \,+\,\\
\label{eq:sigma_a_sigma_matrix_structure}
& \,+\,\alpha{1\over\Omega}a_\sigma
\left(\begin{array}{cc}  \tilde{\Delta}^2\tau_- & \Delta\epsilon\tau_-(\sigma_z-\sigma) \\ 
{\tilde{\Delta}^2\over\Delta}\epsilon(\sigma_z-\sigma)\tau_-  & \epsilon^2(1-\sigma\sigma_z) \end{array}\right) 
\quad,
\end{align}
where $a_\sigma=2\sigma\ln{2}-i\pi$. The transformed Liouvillian $A\tilde{L}_0A$ is given by
(\ref{eq:tilde_L_0_transformed}). Due to the matrix structure of $\tilde{\Sigma}_a(E)$, one 
projector is exactly known (in all orders of perturbation theory, see (\ref{eq:P_st}) and
(\ref{eq:sigma_a_matrix_structure}))
\begin{align}
\label{eq:tilde_P_st_i}
A \tilde{P}^i_{\text{st}} A\,&=\,
\left(\begin{array}{cc} \tau_+ & 0 \\ 0 & 0  \end{array}\right)\quad.
\end{align}
Using usual perturbation theory it is straightforward to calculate the projectors 
$\tilde{P}_\sigma^{\text{st},0,\sigma}$ in zero and first order in  $\alpha$ as
\begin{align}
\label{eq:P_0_i_sigma_app}
A \tilde{P}^{(0)i}_\sigma A\,&=\,{1\over 2}
\left(\begin{array}{cc} 0 & 0 \\ 0 & 1+\sigma\sigma_z \end{array}\right)\quad,\\
\nonumber
A \tilde{P}^{(1)\text{st},0}_\sigma A\,&=\\
\label{eq:P_1_st/0_sigma_app}
&\hspace{-1.5cm}
=\,-{1\over 2}i\pi\alpha{\tilde{\Delta}^2\epsilon\over\Delta\Omega^2}
\left(\begin{array}{cc} 0 & {1\over Z}\tau_-(1+\sigma\sigma_z) \\ 
(1+\sigma\sigma_z)\tau_- & 0 \end{array}\right)\quad,\\
\label{eq:P_1_sigma_sigma_app}
A \tilde{P}^{(1)\sigma}_\sigma A\,&=\,{1\over 4}i\pi\sigma\alpha{\tilde{\Delta}^2\over\Omega^2}
\left(\begin{array}{cc} 0 & 0 \\ 0 & \tau_+-\tau_- \end{array}\right)\quad.
\end{align}
Using $\sum_{j=\text{st},0,\pm}A\tilde{P}_j^iA=1$, we find
\begin{align}
\label{eq:P_0_i_0_app}
A \tilde{P}^{(0)i}_0 A\,&=\,
\left(\begin{array}{cc} \tau_- & 0 \\ 0 & 0  \end{array}\right)\quad,\\
\label{eq:P_1_st/0_0_app}
A \tilde{P}^{(1)\text{st},0}_0 A\,&=\,i\pi\alpha{\tilde{\Delta}^2\epsilon\over\Delta\Omega^2}
\left(\begin{array}{cc} 0 & {1\over Z}\tau_- \\ \tau_- & 0  \end{array}\right)\quad.
\end{align}
This proves (\ref{eq:P_0_i_0}-\ref{eq:P_1_sigma_sigma}) and (\ref{eq:P_st_st}-\ref{eq:P_1_st_0}). 
We note that although degenerate perturbation theory is needed to calculate $\tilde{P}_0^i$ up
to first order in $\alpha$, we do not need any second order terms in $\alpha$ for the Liouvillian
since the projector $\tilde{P}_\text{st}^i$ is exactly known. This is a particular advantage for the
spin boson model.

To derive the formula (\ref{eq:gamma_0_st}) for the eigenvalue $\tilde{\gamma}_0^\text{st}$ of
$\tilde{L}_a(0)$ up to second order in $\alpha$, we relate it to the eigenvalue 
$\tilde{\gamma}_0^0=z_0=-i(\Gamma^{(1)}+\Gamma^{(2)}+O(\alpha^3))$ of $\tilde{L}_a(z_0)$. We first note 
that, due to the matrix structure of $A\tilde{L}_a^{\text{st},0}A$ 
(see (\ref{eq:tilde_L_0_transformed}) and (\ref{eq:sigma_a_0_matrix_structure})), 
the second order contribution to the eigenvalues $\tilde{\gamma}_0^{\text{st},0}$ is not influenced by
the non-diagonal blocks of $A\tilde{L}_a^{\text{st},0}A$ and arises only from the upper left block.
Denoting this block by $(A\tilde{L}_a^{\text{st},0}A)_{11}$ we expand
\begin{align}
\label{eq:relation_sigma_a_z0_0}
\tilde{L}_a(0)_{11}\,=\,\tilde{L}_a(z_0)_{11}\,-\,{d\tilde{\Sigma}_a\over dE}(0)_{11}z_0\,+\,O(\alpha^3)
\quad.
\end{align}
and use (\ref{eq:renormalized_sigma_a}) together with 
${d\over dE}{\cal{F}}_\sigma(0)=1+i\sigma{\pi\over 2}+O(\alpha)$ to get
\begin{align}
\tilde{L}_a(0)_{11}\,=\,\tilde{L}_a(z_0)_{11}\,+\,
2i\Gamma^{(1)}\alpha{\tilde{\Delta}^2\over\Omega^2}\tau_- \,+\,O(\alpha^3)\quad,
\end{align}
such that 
\begin{align}
\label{eq:diff_gamma_0_st}
\tilde{\gamma}_0^{\text{st}}\,=\,z_0 \,+\,2i\Gamma^{(1)}\alpha{\tilde{\Delta}^2\over\Omega^2}\,+\,O(\alpha^3)
\quad,
\end{align}
which proves (\ref{eq:gamma_0_st}).

\end{appendix}

\end{document}